\documentclass[aps,prl,amsmath,amssymb,groupedaddress,10pt,superscriptaddress,floatfix,twocolumn,showkeys,longbibliography,amsart]{revtex4-1} 

\usepackage{array,amsmath,amsfonts,amssymb,dsfont,tabularx,multirow}
\usepackage{hhline}
\usepackage[T1]{fontenc} \usepackage{ae} \usepackage{color} \usepackage{graphicx}
\usepackage{soul} 
\usepackage[dvipsnames]{xcolor}

\usepackage{amssymb}
\usepackage{pifont}
\newcommand{\norm}[1]{\lVert#1\rVert}

\usepackage{makecell}

\usepackage{gensymb} 

\begin{document}

\title{Self-calibrating vector atomic magnetometry through microwave polarization reconstruction}

\author{T.~Thiele}\email{tobias.thiele@colorado.edu} 
\author{Y.~Lin}\thanks{current address: CAS Key Laboratory of Microscale Magnetic Resonance and Department of Modern Physics, University of Science and Technology of China, Hefei 230026, China; Synergetic Innovation Center of Quantum Information and Quantum Physics, University of Science and Technology of China, Hefei 230026, China}
\author{M. O. Brown} 
\author{C. A. Regal} \affiliation{JILA, National Institute of Standards and Technology and University of Colorado, and
Department of Physics, University of Colorado, Boulder, Colorado 80309, USA}

\pacs{}

\renewcommand{\i}{{\mathrm i}} \def\1{\mathchoice{\rm 1\mskip-4.2mu l}{\rm 1\mskip-4.2mu l}{\rm
1\mskip-4.6mu l}{\rm 1\mskip-5.2mu l}} \newcommand{\ket}[1]{|#1\rangle} \newcommand{\tybra}[1]{\langle
#1|} \newcommand{\braket}[2]{\langle #1|#2\rangle} \newcommand{\kebtra}[2]{|#1\rangle\langle#2|}
\newcommand{\opelem}[3]{\langle #1|#2|#3\rangle} \newcommand{\projection}[1]{|#1\rangle\langle#1|}
\newcommand{\scalar}[1]{\langle #1|#1\rangle} \newcommand{\op}[1]{\hat{#1}}
\newcommand{\vect}[1]{\boldsymbol{#1}} \newcommand{\id}{\text{id}}

\begin{abstract}
Atomic magnetometry is one of the most sensitive ways to measure magnetic fields.  We present a method for converting a naturally scalar atomic magnetometer into a vector magnetometer by exploiting the polarization dependence of hyperfine transitions in rubidium atoms. First, we fully determine the polarization ellipse of an applied microwave field using a self-calibrating method, \textit{i.e.}~a method in which the light-atom interaction provides everything required to know the field in an orthogonal laboratory frame. We then measure the direction of an applied static field using the polarization ellipse as a three-dimensional reference defined by Maxwell's equations. Although demonstrated with trapped atoms, this technique could be applied to atomic vapors, or a variety of atom-like systems.

\end{abstract} 		\maketitle

Sensitive magnetometers are increasingly important in both fundamental and technological applications. High accuracy and precision magnetometers are used for dark matter searches and aid in tests of fundamental symmetries. They enable applications ranging from navigation, timekeeping, and geophysical measurement to biological imaging.  A wide array of application-specific requirements has yielded a wide array of magnetometry technologies, drawing on atomic vapors~\cite{Bloom_principles_1962,budker_optical_2007}, nitrogen-vacancy centers~\cite{Rondin_magnetometry_2014}, nuclear magnetic resonance~\cite{Gross_dynamic_2016}, and superconducting quantum-interference devices~\cite{kleiner_superconducting_2004}.

For many magnetometry applications measurement of the scalar field is sufficient, but also knowing the field's full vector description can have important implications, in particular in geosensing~\cite{goetz_remote_1983,reid_magnetic_1990,Lenz_review_1990} and the calibration of precision physics experiments~\cite{Budker_optical_2013}. However, mapping a magnetic field in three-dimensional space with a robust calibration is nontrivial, and presents distinct challenges in different platforms.

Superconducting quantum interference devices (SQUIDS) and Hall or fluxgate sensors are naturally sensitive to a field component perpendicular to, for example, a current loop.  But multiple sensors must be used to measure the field in all three dimensions, and common problems are drifts or uncertainties in the relative directions of the axes~\cite{Camps_numerical_2009,Liu_Novel_2014}.  Solid-state sensors such as nitrogen vacancy (NV) centers in diamond have emerged as a robust and broadband room-temperature platform for magnetic sensing and imaging.  The inherent crystalline structure of NVs provides a natural reference for vector sensing that is actively being developed~\cite{Alegre_polarization_2007,maertz_vector_2010,wang_high_2015,lee_vector_2015,munzhuber_polarization_2017,schloss_simultaneous_2018}.

The most precise magnetometers, which reach sensitivities beyond $\text{fT}/\sqrt{\text{Hz}}$, are atomic magnetometers that consist of many indistinguishable atoms in the vapor phase~\cite{kominis_subfemtotesla_2003}. However, as they are based on Larmor precession, they are scalar sensors, and there is no natural knob for breaking down the total field into components. In the most standard approach to an atomic vector magnetometer, vector addition of an applied static bias field and the field to be measured can be used to extract the unknown field direction~\cite{gravrand_calibration_2001,Vershovskii_fast_2006,seltzer_unshielded_2004}. However, knowledge of the applied bias fields in a orthogonal laboratory frame is limited by the calibration of the external coil set used to apply the bias field. To avoid reliance upon mechanical construction tolerances for calibration, a number of ideas have been developed for atomic vector magnetometers, such as double-resonance magnetometers~\cite{weis_theory_2006,Pustelny_nonlinear_2006,Ingleby_vector_2018}, the use of electromagnetically-induced-transparency effects~\cite{Cox_measurements_2011,Lee_sensitive_1998}, and orthogonal pump beams and effective fields of optical light~\cite{patton_all_2014}.

In this Letter, we introduce a spatial reference for vector atomic magnetometry based upon the three-dimensional structure of a microwave field.  Our work draws on advances in another domain of magnetometry - that of microwave-field measurements and imaging based upon atomic spectroscopy~\cite{affolderbach_imaging_2015,kinoshita_measurement_2016,horsley_frequency_2016}. 
In these techniques, a microwave field can be characterized through dependence of the atomic response on the polarization of the microwave radiation with respect to an applied quantization axis. In our work, we demonstrate a general algorithm for full reconstruction of a microwave polarization ellipse based upon atomic measurements. Importantly, we present how mapping the three-dimensional ellipse is self-calibrating:  Systematics in the direction and strength of applied bias fields, \textit{e.g.}~non-orthogonal field orientations, can be located and corrected based upon the expected atomic response and electro-magnetic field structure. 

Using the reconstructed microwave polarization ellipse as a fundamental reference, we demonstrate atomic vector magnetometry with a multi-level atom. We measure the strength and direction of an applied static magnetic field using only the microwave polarization information and the relative strength of atomic transitions, without the need for rotation of additional static fields.  Our magnetometer can operate in either small field or with an applied reference field.

Our experiments take place using single trapped alkali atoms; while the sensitivity of the experiment undertaken with a few atoms is limited, it enables a proof-of-principle demonstration.   In trapped atom experiments, developing knowledge of applied microwave polarization can be useful for optimization of atomic Rabi rates, characterization of effective magnetic fields in complex trapping potentials~\cite{kien_dynamical_2013,thompson_coherence_2013,schneeweiss_nanofiber_2014,petersen_chiral_2014,goban_atom-light_2014,kaufman_laser-cooling_2015}, and calibration of bias field directions for atomic clocks.  However, in the context of atomic magnetometry, we envision our technique will be most relevant to hot-vapor cells, where one can measure magnetic fields with greater precision and versatility.

\begin{figure}[t]\centering \includegraphics[width=86mm]{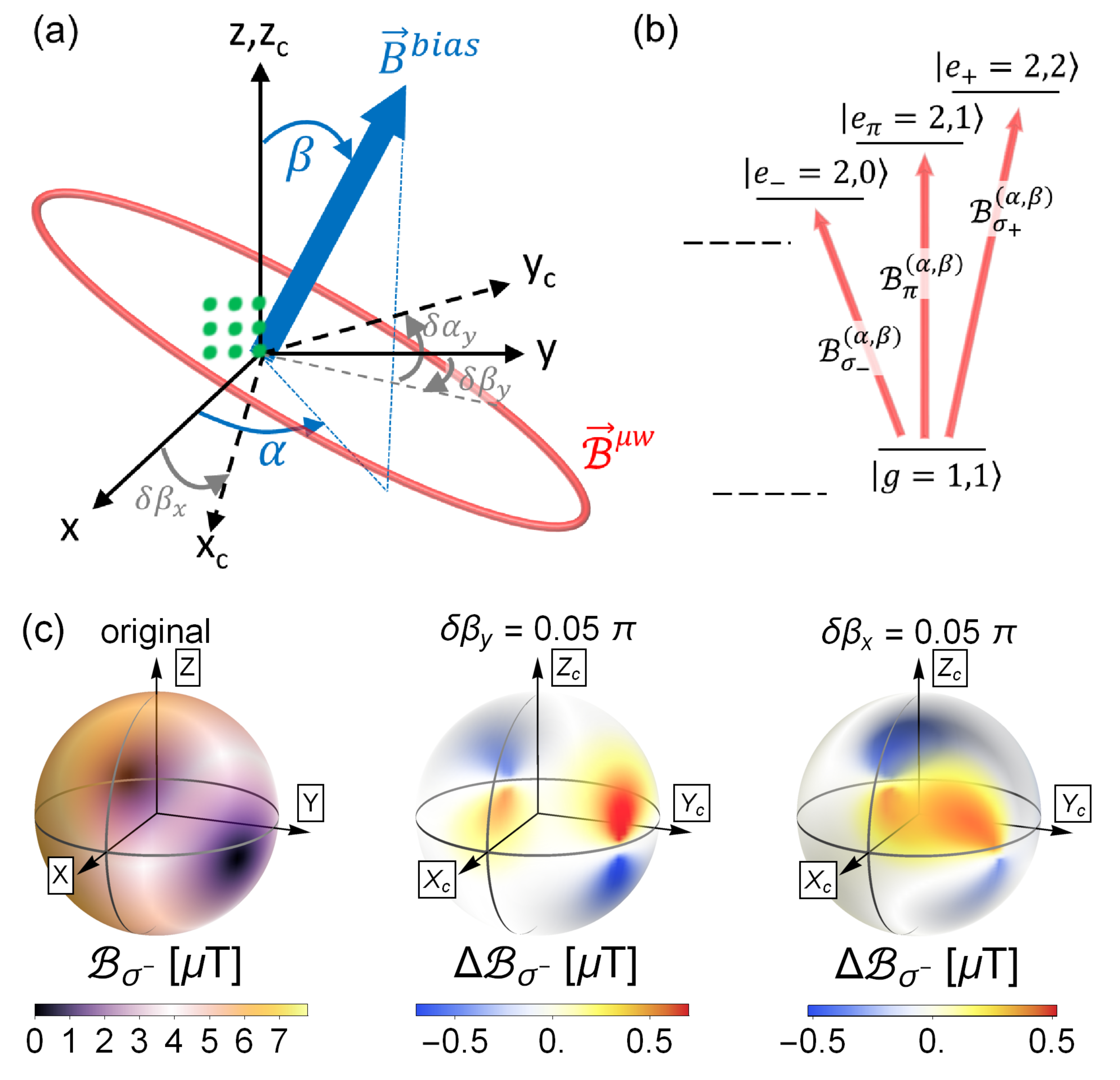}
\caption{(a) Sketch of experimental setup. $^{87}$Rb atoms (green) are trapped in nine optical tweezers and coherently manipulated using microwave radiation (red ellipse). The direction of a magnetic bias field $\vec{B}^\text{bias}$ (blue arrow) can be rotated in any direction given by Euler angles $\alpha$ and $\beta$ defined in the laboratory frame $(\vec{x},\vec{y},\vec{z})$. The bias coils form a slightly non-orthogonal coordinate frame $(\vec{x}_\text{c},\vec{y}_\text{c},\vec{z}_\text{c})$ with angles relative to the laboratory frame of $(\delta\beta_x,\delta\alpha_y,\delta\beta_y)$. (b) Hyperfine transitions (red) between Zeeman-sublevels of $^{87}$Rb. (c) [left panel]  Strength of the $\mathcal{B}_{\sigma_-}$ component of a microwave field for all directions $(\alpha,\beta)$ of $\vec{B}^\text{bias}$ in the laboratory frame for an example polarization ellipse. [center and right panels] Differences in measured and predicted $\mathcal{B}_{\sigma_-}$ for the sample values of $\delta \beta_x$ and $\delta \beta_y$ are indicated.}
\label{fig:Setup}
\end{figure}

We begin by describing our specific experimental setup and atomic multi-level structure, although the procedures we describe apply generally. We use single $^{87}$Rb atoms loaded with $50\%$-probability into a regular $3\times3$ array of $1.7$ $\mu$m-spaced optical tweezers [Fig.~\ref{fig:Setup}(a)]~\cite{kaufman_laser-cooling_2015}. We use four levels of $^{87}$Rb:  A ground-state $\ket{g}=\ket{\text{F}=1,m_\text{F}=1}$ and three excited hyperfine states $\ket{e_+}=\ket{2,2}$, $\ket{e_\pi}=\ket{2,1}$ and $\ket{e_-}=\ket{2,0}$ states.  

We start the experiment by initializing the atoms in $\ket{g}$. During the experiment, we drive to the excited states using $\sigma_\pm$ and $\pi$-polarized light components of a $6.834$-GHz-microwave field ($\lambda\approx44~\text{mm}$) of magnitude $|\vec{\mathcal{B}}^{\mu\text{w}}|_\text{max}\equiv\norm{\vec{\mathcal{B}}^{\mu\text{w}}}\approx 7.8~\mu$T 
[Fig.~\ref{fig:Setup}(b)]. 

The transitions are split by a $300$-$\mu$T-strong static magnetic bias field $\vec{B}^\text{bias}$. The maximal splitting between the three transitions  ($\sim 4.2$ MHz) is $<0.1\%$ of the microwave frequency, and hence spatial field differences are irrelevant when resonant with a transition. $\vec{B}^\text{bias}$ is controlled by three coil pairs in near-Helmholtz configurations that define a coil-frame $\mathcal{C}=(\vec{x}_\text{c},\vec{y}_\text{c},\vec{z}_\text{c})$ that importantly is not necessarily orthogonal. An orthonormal laboratory frame $\mathcal{L}=(\vec{x},\vec{y},\vec{z})$ is chosen s.t. $\vec{z}$ [Euler angles: $(\alpha,\beta)=(0,0)$] is oriented along $\vec{z}_\text{c}$, and $\vec{x}_\text{c}$ and $\vec{y}_\text{c}$ point in directions given by angles $(0,\pi/2+\delta \beta_x)$, and $(\pi/2+\delta \alpha_y,\pi/2+\delta \beta_y)$, respectively [Fig.~\ref{fig:Setup}(a)].

We then determine the full polarization ellipse ($PE$) of the magnetic component of our microwave excitation field, expressed in $\mathcal{L}$ as:
\begin{equation}
\vec{\mathcal{B}}^{\mu\text{w}}=\sum_{\substack{j\in \\ \{x,y,z\}}}\frac{1}{2} \mathcal{B}_{j}e^{-i(\phi_{j}+\omega t)}\vec{e}_{j}+\text{c.c.} 
\label{eq:linear}
\end{equation}
$\vec{\mathcal{B}}^{\mu\text{w}}$ traces the $PE$ determined by $5$ independent parameters: $3$ fields $(\mathcal{B}_x,\mathcal{B}_y,\mathcal{B}_z)$ and two relative phases ($\phi_x,\phi_y$), where $\phi_z=0$ without loss of generality \cite{koepsell_measuring_2017}.

The quantization axis of our atoms is defined along $\vec{B}^\text{bias}$, and is always well-defined because $\norm{\vec{\mathcal{B}}^{\mu\text{w}}} \ll |\vec{B}^\text{bias}|$. The microwave field amplitudes $\mathcal{B}_{\pm\sigma}$ and $\mathcal{B}_\pi$ that drive the $\Delta m_\text{F}=\pm1$ and $\Delta m_\text{F}=0$ atomic transitions, respectively, are strongly dependent on the direction of $\vec{B}^\text{bias}$ and hence denoted as $\mathcal{B}_i^{(\alpha,\beta)}$ for $i\in(\sigma_\pm,\pi)$ hereafter. The $\mathcal{B}_i^{(\alpha,\beta)}$ are related to the $5$ polarization-ellipse parameters and the direction of the bias field $(\alpha,\beta)$ by~\cite{koepsell_measuring_2017}:
\begin{subequations}
	\label{eq:amplitude_correspondence}
	\begin{align}
	\left(\mathcal{B}_{\pi}^{(\alpha,\beta)}\right)^{2}=&
    \mathcal{B}_{z}^{2} \cos^2(\beta) \nonumber \\
    &+\left( \mathcal{B}_{x}^{2} \cos^2(\alpha) +  \mathcal{B}_{y}^{2} \sin^2(\alpha)\right)\sin^2(\beta) \nonumber \\
	&+ \mathcal{B}_{z} \mathcal{B}_{x}\sin(2\beta)\cos(\alpha)\cos(\phi_{x})\nonumber \\
	&+ \mathcal{B}_{z} \mathcal{B}_{y}\sin(2\beta)\sin(\alpha)\cos(\phi_{y}) \nonumber \\
	&+ \mathcal{B}_{y} \mathcal{B}_{x}\sin(2\alpha)\sin^2(\beta)\cos(\phi_{x}-\phi_{y}) \\
	\left(\mathcal{B}_{\pm\sigma}^{(\alpha,\beta)}\right)^{2}=& \frac{1}{2}\left(\sum_{\substack{j\in\\  \{x,y,z\}}}ˆ{} \mathcal{B}_{j}^2- \left[\mathcal{B}_{\pi}^{(\alpha,\beta)}\right]^{2}\right) \nonumber \\
	&\pm  \mathcal{B}_{x} \mathcal{B}_{y}\cos(\beta)\sin(\phi_{x}-\phi_{y}) \nonumber \\
	&\mp  \mathcal{B}_{x} \mathcal{B}_{z}\sin(\alpha)\sin(\beta)\sin(\phi_{x}) \nonumber \\
	&\pm  \mathcal{B}_{y} \mathcal{B}_{z}\cos(\alpha)\sin(\beta)\sin(\phi_{y}).
	\end{align}
\end{subequations}
Therefore, to recreate the full PE, $5$ independently chosen measurements of ($\mathcal{B}_{\pm\sigma}^{(\alpha,\beta)}, \mathcal{B}_\pi^{(\alpha,\beta)}$) are enough to solve for the $5$ unknown ellipse parameters~\cite{koepsell_measuring_2017}.  These measurements can either vary any combination of $(\alpha,\beta)$ or the atomic transition used.

We now discuss how we avoid systematic errors that may enter this procedure. First, we measure $\mathcal{B}_{i}^{(\alpha,\beta)}$ by choosing (pulsed) coherent population transfer~\cite{Thiele_supplementarymicrowave_2018,affolderbach_imaging_2015,horsley_frequency_2016,fan_sub_2014,thiele_imaging_2015,koepsell_measuring_2017} to accurately extract $\mathcal{B}_{i}^{(\alpha,\beta)}$ from the corresponding (measured) Rabi frequency $\Omega_i^{(\alpha,\beta)}=\mu_i \mathcal{B}_i^{(\alpha,\beta)}\big/\hbar$.
By referencing $\Omega_i^{(\alpha,\beta)}$ to a frequency standard, $\mathcal{B}_{i}^{(\alpha,\beta)}$ can be determined absolutely when calculating the magnetic transition dipole moments $\mu_i$ from basic assumptions~\cite{Thiele_supplementarymicrowave_2018}. 

However, systematic errors can also enter through discrepancies of the intended and actual applied direction ($\alpha,\beta$) of $\vec{B}^{\text{bias}}$. When performing any directional measurement with an atom(-like) system this is a general limitation which, so far, is typically addressed by relying on the quality of an externally-calibrated system. Let us label the potentially unknown systematic errors by $N^\text{u}$ parameters $\{U_j\}$, $j\in[1,N^\text{u}]$ - these will generally modify the value and direction of $\vec{B}^{\text{bias}}$. The $\{U_j\}$ can be self-calibrated by performing $\geq N^\text{u}$ additional measurements and then solving the system of $\geq N^\text{u}+5$ equations for $(\mathcal{B}_x,\mathcal{B}_y,\mathcal{B}_z,\phi_x,\phi_y;\{U_{j}\})$. 

In our experiment the parameters $\{U_j\}$ are: orientations of the coil pairs $(U_1,U_2,U_3)=(\delta \beta_x,\delta \beta_y,\delta \alpha_y)$ forming coil frame $\mathcal{C}$ [Fig.~\ref{fig:Setup}(a)], components of a (stray) magnetic field $(U_4,U_5,U_6)=(B_x^\text{s},B_y^\text{s},B_z^\text{s})$ due to imperfect nulling of the ambient magnetic fields, and calibration errors in the applied bias field $(U_7,U_8,U_9)=(|\vec{B}^\text{bias}|,\epsilon_x,\epsilon_y)$ when parameterizing $\vec{B}^\text{bias}=|\vec{B}^\text{bias}|(\epsilon_x b_x,\epsilon_y b_y,b_z)$ using $\sum_{i\in (x,y,z)} b_\text{i}^2=1$~\cite{Thiele_supplementarymicrowave_2018}. For completeness, this set of parameters also includes the magnitude of $\vec{B}^\text{bias}$ that will be determined from Zeeman shifts, which is irrelevant for determining a PE, but defines a common scaling factor for all static fields. We therefore need $\geq 14$ measurements to self-calibrate our magnetometer and the bias-field strength.

Our self-calibration of the $\{U_j\}$ stems from the structure of the microwave light dictated by Maxwell's equations. To illustrate the key idea, we consider two simple examples where either $U_1=\delta\beta_x$ or $U_2=\delta\beta_y$ are unknown. Assume all of the $N^\text{u}$ measurements are performed on the $\mathcal{B}_{\sigma_-}^{(\alpha,\beta)}$-component of an elliptically polarized microwave field.  The expected values of $\mathcal{B}_{\sigma_-}$ for an example PE are shown in the lab frame $\mathcal{L}$ as a function of $(\alpha,\beta)$ [Fig.~\ref{fig:Setup}(c)]. If the coil frame deviates from the lab frame such that $(\beta_x,\beta_y)=(0,0.05\pi)$ the functional form of $\mathcal{B}_{\sigma_-}^{(\alpha,\beta)}=\mathcal{B}_{\sigma_-}(\alpha,\beta;\beta_y)$ will deviate from the expectation in $\mathcal{L}$ with a specific pattern $\Delta \mathcal{B}_{\sigma_-}$ [center panel in Fig.~\ref{fig:Setup}(c)].  This field pattern cannot be reproduced by allowed microwave ellipses, and is distinctly connected to the unknown parameter. Importantly, a different pattern is associated with $(\beta_x,\beta_y)=(0.05\pi,0)$ [right panel in Fig.~\ref{fig:Setup}(c)].  Suitably chosen measurements on the sphere can hence lead to full differentiation and absolute characterization of the unknowns $\{U_j\}$.  

\begin{figure}[t] \centering \includegraphics[width=86mm]{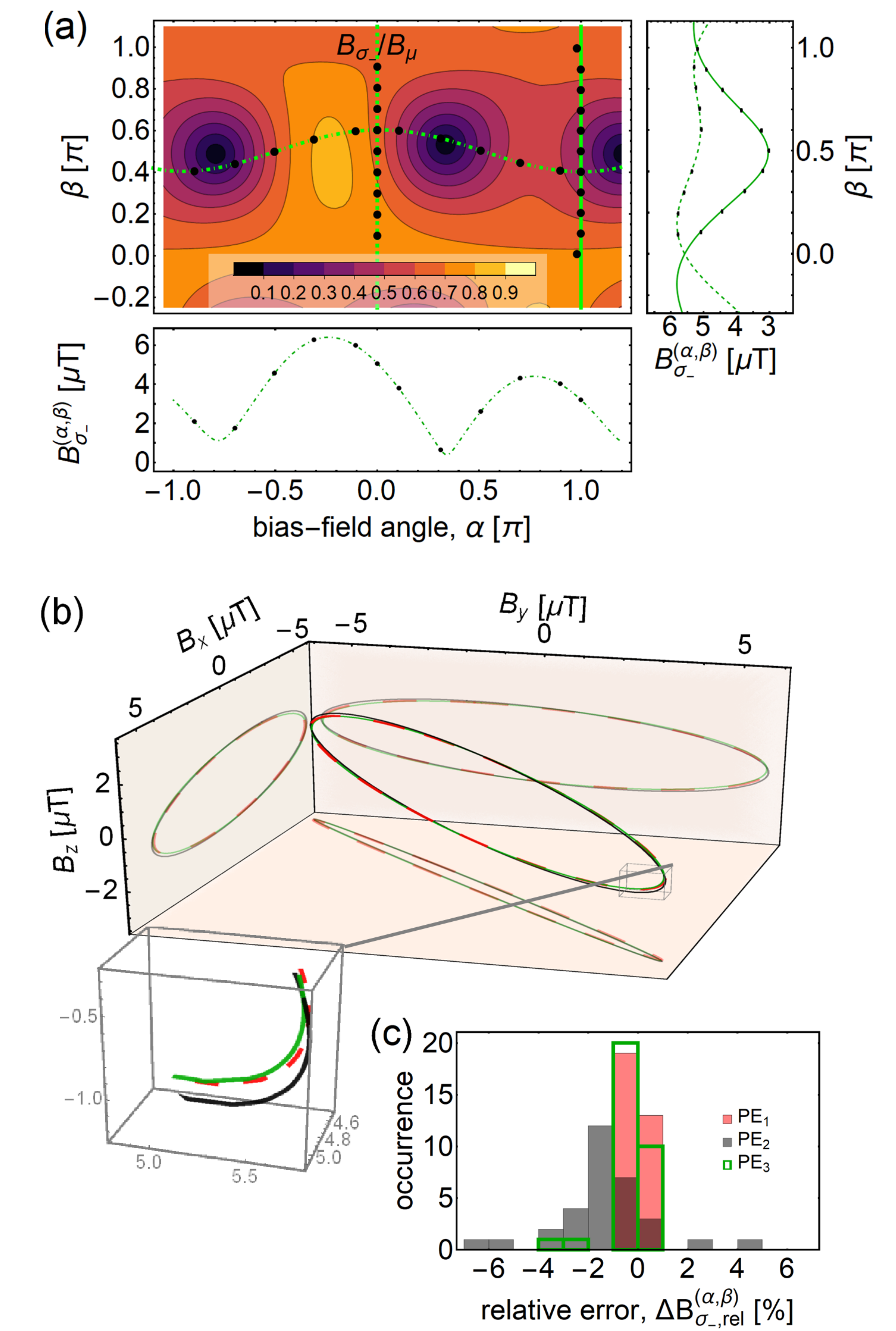}
\caption{(a) (2D-plot) Predicted magnetic field component $\mathcal{B}_{\sigma_-}^{(\alpha,\beta)}$ using parameters from the polarization ellipse $PE_1$, as a function of the  bias-field direction $(\alpha,\beta)$. (small panels) Measured magnetic microwave field component (black data) and predictions along the solid, dashed, and dot-dashed cuts in the 2D-plot, respectively. The measurement uncertainties are smaller than the datapoints. (b) Polarization ellipses $PE_1$(dashed, red), $PE_2$(black) and $PE_3$(green), respectively. $PE_1$ is the reference ellipse from the $30$ calibration measurements of $\mathcal{B}_{\sigma_-}^{(\alpha,\beta)}$ from panel (a). Comparison of ellipses $PE_2$ ($PE_3$) assess our protocol using $5$ of $6$ measurements~\cite{Thiele_supplementarymicrowave_2018} without (with) taking into account calibrated $\{U_j\}$, see text. (c) Overlayed histograms of the relative errors $\Delta \mathcal{B}_{\sigma_-,\text{rel.}}^{(\alpha,\beta)}$ of all polarization ellipses and for all measurements displayed in panel (a). The colorcoding is the same as for panel (b).}
\label{fig:check}
\end{figure}

To calibrate the $9$ $\{U_j\}$ in our experiment, we measure $\mathcal{B}_{\sigma_-}^{(\alpha,\beta)}$ (together with the Zeeman-shift of the $\sigma_-$-transition) for $28$ different directions $(\alpha,\beta)$ [black points in Fig.~\ref{fig:check}(a)]. Using quadratic minimization~\cite{Thiele_supplementarymicrowave_2018} and Eq.~(\ref{eq:amplitude_correspondence}b), these are enough measurements to determine $(\delta\beta_x,\delta\alpha_y,\delta\beta_y)=(1.3\,\text{mrad},10.9\,\text{mrad},5.4\,\text{mrad}),~(B_x^\text{s},B_y^\text{s},B_z^\text{s})=(-6.05\,\mu\text{T},0.14\,\mu\text{T},-1.12\,\mu\text{T}),~|\vec{B}^\text{bias}| = 302.0~\mu\text{T},$ $(\epsilon_x,\epsilon_y) = (1.001,0.989)$, and a polarization ellipse we refer to as $PE_1$: $(\mathcal{B}_x,\mathcal{B}_y,\mathcal{B}_z,\phi_x,\phi_y)=(5.023(5)\,\mu\text{T},~5.757(4)\,\mu\text{T},~1.600(4)\,\mu\text{T},~-1.941(4),$ $-1.873(4))$ [red, dashed ellipse in Fig.~\ref{fig:check}(b)]. We cannot easily extract uncertainties for the $\{U_j\}$, but estimations have shown that we need to vary a $U_j$ by $\sim20\%$ to change $PE_1$ by its uncertainty of $\lesssim0.2\%$. Furthermore, the result is within expectation of experimentally-defined parameters in our setup~\cite{Thiele_supplementarymicrowave_2018}; for example  the angles measured are consistent with machining tolerances of the coil mounts.  The measured Zeeman shift allows us to determine the absolute value of $|\vec{B}^\text{bias}|$ that we find has a consistent dependence on $(\alpha,\beta)$~\cite{Thiele_supplementarymicrowave_2018}. 

Figure~\ref{fig:check}(a) predicts $\mathcal{B}_{\sigma_-}^{(\alpha,\beta)}$ for all directions of $\vec{B}^\text{bias}$ using Eq.~(\ref{eq:amplitude_correspondence}b), and could be used to optimize the atom-light coupling strength by choosing a suitable $\vec{B}^\text{bias}$. Furthermore, quantitative comparison with experiment provides a measure of how well $\{U_j\}$ and the ellipse parameters have been determined. For this, we investigate the relative errors $\Delta \mathcal{B}_{\sigma_-,\text{rel.}}^{(\alpha,\beta)}$ of all measured $\mathcal{B}_{\sigma_-}^{(\alpha,\beta)}$ and their predicted values [see red histogram Fig.~\ref{fig:check}(c)]. Not surprisingly, we find that all $\Delta \mathcal{B}_{\sigma_-,\text{rel.}}^{(\alpha,\beta)}$ are distributed around $0$, as they have been used to determine $PE_1$. The width of the distribution is consistent with the $\vec{\mathcal{B}}_{\sigma_-}^{(\alpha,\beta)}$ measurement uncertainty of $\sim 0.1\%$, and a microwave-amplitude drift of $\lesssim \pm 1\%$ occurring on timescales of several minutes~\cite{Thiele_supplementarymicrowave_2018}. This drift could be stabilized in future experiments.

To elucidate experimentally self-calibration we determine a new polarization ellipse from a completely independent set of measurements: $\mathcal{B}_{\sigma_+}^{(0,0)},~\mathcal{B}_{\sigma_-}^{(0,0)},~\mathcal{B}_{\sigma_+}^{(0,\pi/2+\delta\beta_x)},~\mathcal{B}_\pi^{(0,\pi/2+\delta\beta_x)}$, and $\mathcal{B}_{\sigma_-}^{(0,\pi/2+\delta\beta_x)}$, for which we rotate our bias field from the $\vec{z}_\text{c}$ to the $\vec{x}_\text{c}$ direction in the coil frame~\cite{Thiele_supplementarymicrowave_2018}. Note, before we measured a single polarization for many directions, and now we measure all three polarization components for only two directions, \textit{i.e.} we only rotate the bias field in a single plane \textit{once}. Nonetheless, using Eqs.~(\ref{eq:amplitude_correspondence}), we determine the microwave field in all three dimensions. 

First, we assume (wrongly) that all $U_j=0$ (except $U_7=|\vec{B}^\text{bias}|=300~\mu$T), \textit{i.e.}~our coils are perfectly orthogonal, calibrated, and no stray magnetic fields are present. From this, we obtain $PE_2$: $(5.126(1)\,\mu\text{T},~5.786(1)\,\mu\text{T},~1.6537(7)\,\mu\text{T},~-1.970(3),$ $-1.910(2))$, depicted in black in Fig.~\ref{fig:check}(b). The shape and orientation of $PE_2$ agrees with $PE_1$ despite the different ways they were determined; but they are not identical. This is also captured in the corresponding distribution of $\Delta \mathcal{B}_{\sigma_-,\text{rel.}}^{(\alpha,\beta)}$ of $PE_2$ [black histogram in Fig.~\ref{fig:check}(c)]. Its center is slightly offset from $0$, and indicates relative errors larger than $5\%$. These are larger than the slow drifts of our microwave source and the measurement uncertainties~\cite{Thiele_supplementarymicrowave_2018}.

However, using the same $5$ measurements as for $PE_2$ now taking into account the correctly calibrated $\{U_j\}$, and a microwave drift correction of 1\%~\cite{Thiele_supplementarymicrowave_2018}, we obtain $PE_3$: $(5.070(1)\,\mu\text{T},~5.743(1)\,\mu\text{T}, ~1.6018(6)\,\mu\text{T},$ $-1.910(3),~-1.844(3))$ [green ellipse Fig.~\ref{fig:check}(b)]. This polarization ellipse predicts again the measurements correctly, \textit{i.e.} the width of the $\Delta \mathcal{B}_{\sigma_-,\text{rel.}}^{(\alpha,\beta)}$-distribution is consistent with our microwave drifts~\cite{Thiele_supplementarymicrowave_2018}.  

With the microwave field as a static, well-calibrated reference in the laboratory frame $\mathcal{L}$, we now use the atoms to vectorially resolve a set of $3$ intentionally applied static probe fields $\vec{B}^\text{p}$ in $\mathcal{L}$. The procedure: combine any scalar atomic measurement of $|\vec{B}^\text{p}|$ with two Rabi-rate measurements to solve for its orientation. Specifically, the $\sim200~\mu$T-strong-probe magnetic fields $\vec{B}_j^\text{p}~[j=(x_\text{c},y_\text{c},z_\text{c})$] are sequentially applied along $\vec{x}_\text{c}$, $\vec{y}_\text{c}$, and $\vec{z}_\text{c}$, respectively, in addition to a reference field $\{|\vec{B}^\text{ref}|,\,(\alpha,\beta)^\text{ref}\} \approx \{300~\mu\text{T}, (0.1\pi,0.6\pi)\}$. Then we measure the total magnetic bias field $\vec{B}_j^\text{m}=\vec{B}_j^\text{p}+\vec{B}^\text{ref}$ (and, for completeness $\vec{B}^\text{ref}$) and determine $\vec{B}_j^\text{p}$ by subtraction of $\vec{B}^\text{ref}$. The use of a reference field $\vec{B}^\text{ref}\neq0$ is not necessary, but can be useful (see supplement~\cite{Thiele_supplementarymicrowave_2018}). All magnetic field \textit{magnitudes} are found from the mean of the atomic Zeeman-shifts of all three available transitions. The \textit{directions} $(\alpha,\beta)_j$ of $\vec{B}_j^\text{m}$ and $\vec{B}^\text{ref}$ are determined by measuring $\mathcal{B}_{\sigma_-}^{(\alpha,\beta)_j}$ and $\mathcal{B}_\pi^{(\alpha,\beta)_j}$ and then use Eq.~(\ref{eq:amplitude_correspondence}) to solve for $(\alpha,\beta)_j$ via quadratic minimization~\cite{Thiele_supplementarymicrowave_2018}.  We also measure $\mathcal{B}_{\sigma_+}^{(\alpha,\beta)_j}$, but use this for keeping track of drifts in the amplitude of our applied microwave field. 

\begin{figure}[t] \centering \includegraphics[width=86mm]{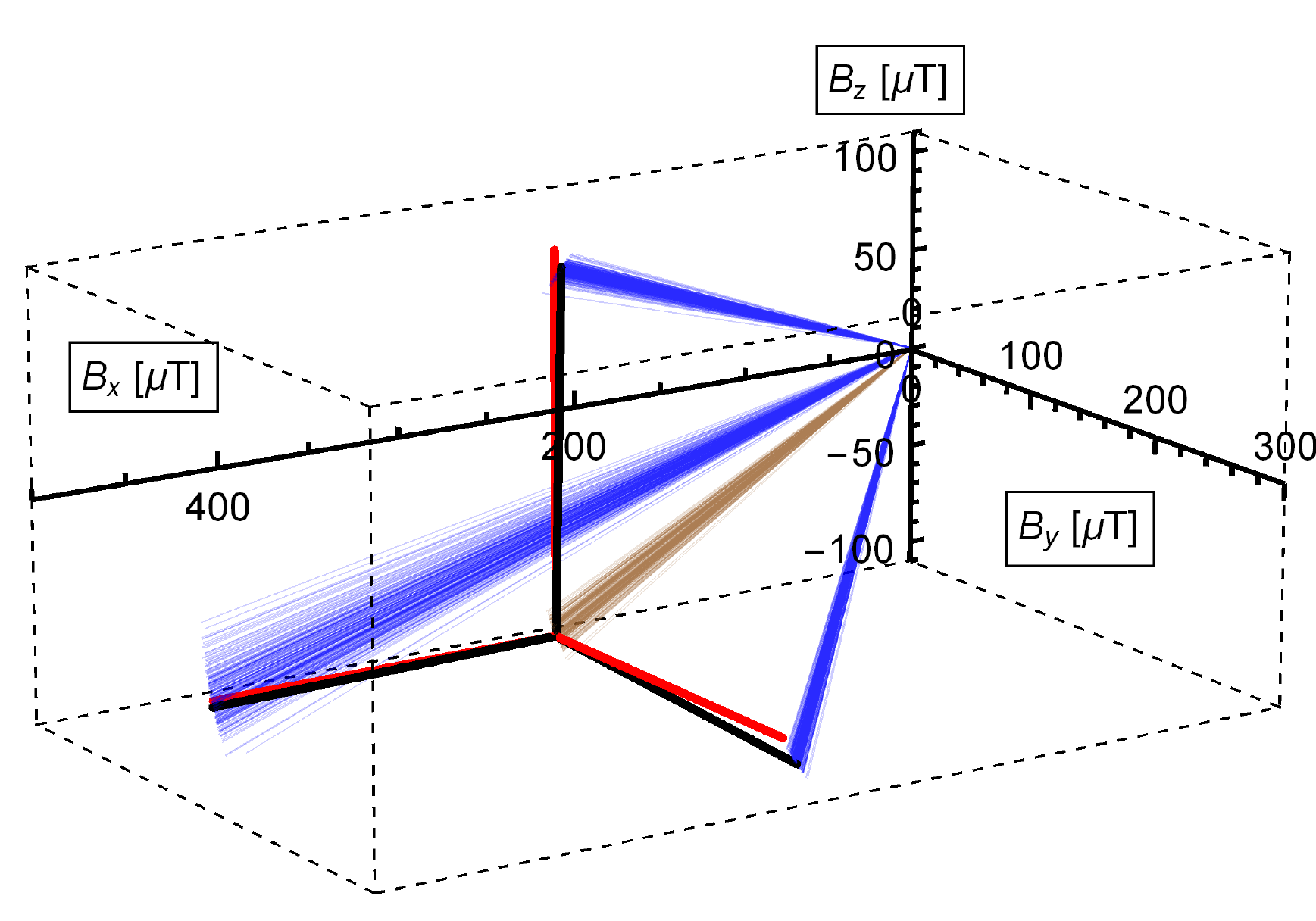}
\caption{Vector magnetometry in an orthonormal laboratory frame $\mathcal{L}$: Targeted (red) applied magnetic field vectors $\vec{B}_j^\text{p}$, $j\in(x_\text{c},y_\text{c},z_\text{c})$, can be reconstructed (black) as the difference between all measured vectors $\vec{B}_j^\text{m}$ (dark blue lines) and an initially applied reference field $\vec{B}^\text{ref}$ (brown lines). The multiple lines for all displayed measured fields indicate the range of measurement errors, determined by bootstrapping the error with $200$ trials.
}
\label{fig:VectorMagnetometer}
\end{figure}

We find $\{|\vec{B}^\text{ref}|,\,[\alpha,\beta]^\text{ref}\}=\{296.6(2)\,\mu\text{T},\,[0.103(1)\pi,$ $0.588(6)\pi]\}$ (brown lines). Sequential application of the $3$ probe fields $\vec{B}_j^\text{p}$ results in total measured fields $\{|\vec{B}_j^\text{m}|,$ $[\alpha,\beta]^\text{m}_j\} = \{476.1(8)\,\mu\text{T},~[0.0636(6)\pi,0.554(6)\pi]\}$, $\{301.5(3)\,\mu\text{T},$ $[0.108(2)\pi,0.396(4)\pi]\}$, and $\{399.2(5)\,\mu\text{T},$ $[0.260(1)\pi,0.577(2)\pi]\}$ [solid blue lines in Figure ~\ref{fig:VectorMagnetometer}]. The multiple lines represent measurement uncertainties from bootstrapping the error with 200 trials. The difference in the uncertainties for these measurements is determined by the precision with which we measure the $\Omega_{i}^{(\alpha,\beta)}$, and by its transfer function to the field direction. This transfer function causes the uncertainties to be very asymmetric (aspect ratios up to $30$) and is ultimately linked to the choice of $\vec{B}^\text{ref}$~\cite{Thiele_supplementarymicrowave_2018}.

 We determine the mean probe fields $\vec{B}_j^\text{p}=\vec{B}_j^\text{m}-\vec{B}^\text{ref}$ 
(black) as the difference between the blue ($\vec{B}_j^\text{m}$) and the brown ($\vec{B}^\text{ref}$) vectors. The probe fields that we expect (red lines) point in $\vec{x}_\text{c}$, $\vec{y}_\text{c}$ and $\vec{z}_\text{c}$-direction from $\vec{B}^\text{ref}$, and deviate from
$\vec{B}_j^\text{m}-\vec{B}^\text{ref}$ by $16$ mrad, $79$ mrad and $23$ mrad, respectively. For the $\vec{y}_\text{c}$ and $\vec{z}_\text{c}$-direction, these values are just outside the confidence interval we expect based on the measurement uncertainties, but can be fully explained when including the slow drift of the microwave field strength reported earlier~\cite{Thiele_supplementarymicrowave_2018}. Over the course of this set of measurements this drift was found to be $\leq \pm 0.5\%$ as inferred from all the measured magnitudes $7.841(8)~\mu\text{T},\,7.840(6)~\mu\text{T},\, 7.789(5)~\mu\text{T}$, and $7.815(7)~\mu\text{T}$, respectively. 
The magnetometer can operate with reference or without (\textit{reference-free}) $\vec{B}^\text{ref}$. In contrast to the \textit{reference-free} mode, in the \textit{reference-mode} the uncertainty of $\vec{B}^\text{ref}$ adds to the uncertainty of our measurement of $\vec{B}_j^\text{p}$, but makes the measurement of $\vec{B}^\text{p}$ independent of common-mode background fields, a property specifically useful if $\vec{B}^\text{p}$ is time-dependent. Furthermore, $\vec{B}^\text{ref}\neq\vec{0}$ enables control over the precision of the vector magnetometer, and could even allow one to "squeeze" the measured variance in certain directions~\cite{Thiele_supplementarymicrowave_2018}. Lastly, in \textit{reference-mode}, one can avoid the complication of finding up to $4$ field solutions as a result of the measurements combined with the constraints of the trigonometric Eqs.~(\ref{eq:amplitude_correspondence})~\cite{Thiele_supplementarymicrowave_2018}. In addition to this, we discuss in the supplement other strategies involving test-fields to determine the correct solution in the \textit{reference-free mode} ($|\vec{B}^\text{p}|\gg|\vec{B}^\text{ref}|\approx0$).

Looking forward, we envision reconstruction of the microwave polarization ellipse to be a general-purpose reference for not only magnetic fields, but other excitation fields (e.g~optical fields or electric components).  Further assessment is required to understand the potential for precision and sensitivity. Different measurement protocols will also need to be developed to transition these ideas to scalable and more-sensitive atomic vapor cells; coherent population transport was a robust way for us to measure microwave field strengths, but there are a variety of ways to determine the Rabi rates, \textit{e.g.}~by spectroscopic means. As with a variety of atomic sensors, this platform for vector magnetometry is compatible with futuristic quantum-enhancement, as established with cold atoms~\cite{Degen_quantum_2017,Muessel_Scalable_2014}. 

\section{Acknowledgments}
We thank M. Knill for his help with bootstrapping analysis. We thank Svenja Knappe, Philip Treutlein, Ronald Walsworth, Adam Kaufman for helpful discussions, and Chris Kiehl for technical assistance. T.T. acknowledges funding from SNF under project number: P2EZP2\_172208. We acknowledge funding from NSF Grant No. PHYS 1734006, ONR Grant No. N00014-17-1-2245, AFOSR-MURI Grant No. FA9550-16-1-0323, and the David and Lucile Packard Foundation.  M. O. B. acknowledges support from an NDSEG Fellowship.

\end{document}


\title{Supplementary information for 'Self-calibrating vector atomic magnetometry through microwave polarization reconstruction'}

\author{T.~Thiele}\email{tobias.thiele@colorado.edu} 
\author{Y.~Lin}\thanks{current address: CAS Key Laboratory of Microscale Magnetic Resonance and Department of Modern Physics, University of Science and Technology of China, Hefei 230026, China; Synergetic Innovation Center of Quantum Information and Quantum Physics, University of Science and Technology of China, Hefei 230026, China}
\author{M. O. Brown} 
\author{C. A. Regal} \affiliation{JILA, National Institute of Standards and Technology and University of Colorado, and
Department of Physics, University of Colorado, Boulder, Colorado 80309, USA}

\maketitle
\tableofcontents


\section{Experimental setup and measurement procedure}
\label{sec:experimental setup}
Our experiments take place with single $^{87}$Rb atoms loaded into a $3$-by-$3$-square array of optical tweezers. The tweezers have a fixed, minimal trap-spacing of $\sim 1.7~\mu$m in the $(y,z)$-plane [Fig.1(a) in main text~\cite{Thiele_Self_2018}] and are formed by deflecting a single 852-nm-laser beam propagating in $x$-direction using two acousto-optic-modulators~\cite{kaufman_cooling_2012,shih_nondestructive_2013,le_kien_dynamical_2013}. Each trap is gaussian-shaped with a $1/e^2$-radius of $\sim0.71~\mu$m in the radial, \emph{i.e.} $(y,z)$-direction, and has an adjustable trap-depth. Given the spacing and radius of the individual traps, atoms in different traps can be considered independent for all measurements reported. The strength of the bias field $\vec{B}^\text{bias}$ we use to define the quantization axis and to split the hyperfine transitions is created using $3$ pairs of Helmholtz coils. The strength of $\vec{B}^\text{bias}$ can be varied and its direction ($\alpha,\beta$) can be rotated within $2$ ms.

\subsection{Measurement procedure}
\label{sec:measurement procedure}

Every experimental cycle consists of three phases; \emph{initialization}, \emph{experiment} and \emph{detection}.

During \emph{initialization}, the traps are loaded with a probability of $\sim 50\%$ from a dilute optical molasses using polarization-gradient cooling (PGC). To determine which of the $9$ traps are loaded, fluorescence photons of the trapped atoms are collected on a CCD camera during
a second PG-cooling phase (PGC-image). Finally, for $\normbvec{\vec{B}}^\text{bias}\sim 300~\mu$T and $(\alpha,\beta)=(0,0)$, we prepare the atoms in the $\ket{g}=\ket{\text{F}=1,m_\text{F}=1}$ hyperfine ground state by first optically pumping the population to $\ket{2,2}$ ($>99~\%$ fidelity) following a $11.7~\mu$s-long microwave pulse of $\sim 6.834$ GHz ($\sim 95~\%$ efficiency).

During the \emph{experiment} phase, $\vec{B}^\text{bias}$ is rotated into the direction ($\alpha,\beta$) and the trap-depths are lowered to $\sim0.018$ mK in $3$ ms, in order to minimize any trap-effects such as Stark or vector shifts~\cite{petersen_chiral_2014,kien_dynamical_2013,schneeweiss_nanofiber_2014,thompson_coherence_2013,kaufman_laser-cooling_2015}. We then apply a microwave square-pulse of
variable length $\Delta t$, resonant with either of the $\ket{g}\rightarrow(\ket{e_-},\ket{e_\pi},\ket{e_+}=(\ket{2,0},\ket{2,1},\ket{2,2})$ transition during which the atoms undergo coherent Rabi oscillations [Fig.~\ref{fig:Spectrum}(b)]. 

The population in $|g\rangle$ is \emph{detected} by pushing atoms in the excited states out of the traps using resonant light of $\sim 780$ nm at the $5 ^2S_{1/2}|F=2\rangle \rightarrow 5 ^2P_{3/2}|F'=3\rangle$ transition. The atoms in the $\ket{g}$-state remaining in the traps are then revealed using a second PGC-image. For each trap the population in $\ket{g}$ is then given by its survival; that is, for $\sim30$ identical repetitions of the experiment, the average number that an atom is observed in the second image if an atom was detected in the first image.

During the experiment cycle we observe background loss of about $\sim 5\%$ (black area in Fig.~\ref{fig:Spectrum}(b)), and an additional $\sim12\%$ (blue area in Fig.~\ref{fig:Spectrum}(b)) due to the temperature of the atoms compared to the shallow trap depth.

\subsection{Microwave setup and long-term fluctuations of the microwave field strength}
\label{sec:microwavedrift}

In our experiment, square microwave pulses are generated using a series of $2$ switches (forward isolation: $\sim 46$ dB) to shape the $\sim6.8$-GHz-microwave generated from a signal generator (Rohde\&Schwarz SMF100A). The pulses are then amplified using a water-cooled amplifier ($\geq 40$ dB) and directed onto the glass vacuum cell using a sawed-off rectangular waveguide. 

For the measurements for $PE_2$ and $PE_3$, and static vector magnetometry~\cite{Thiele_Self_2018}, we consistently extracted all $3$ components $(\pm\sigma,\pi)$ of the microwave field for a given $\vec{B}^\text{bias}$. With this information we can calculate the total microwave strength at the atoms using $\norm{\vec{\mathcal{B}}^{{\mu\text{w}}}}=\sqrt{\sum_{i\in(\pm\sigma,\pi)}\left(\mathcal{B}_i^{(\alpha,\beta)}\right)^2}$~\cite{koepsell_measuring_2017}, which is independent of the bias-field direction $(\alpha,\beta)$. 

However, as mentioned in the main article~\cite{Thiele_Self_2018}, we observe differences in strength between the measurements (Tab.~\ref{tab:intensities}), which we identify as slow drifts over the course of several minutes.
\begin{table}[t]
	\centering
  	\begin{tabular}{|l||c|c|}
	\hline
\boldmath$[\alpha,\beta]$ & \boldmath$\norm{\vec{\mathcal{B}}^{\mu\text{w}}}$ [$\mu$T]& \boldmath$\Delta \mathcal{B}^{{\mu\text{w}}}$ [\%]\\ 
\hhline{|=#=|=|}
$[0,0]$ & 7.814(2) & -0.26 \\
\hline 
$[0,0.5015\pi]$ & 7.906(1)& 0.92 \\
\hline
$[0.102(1)\pi, 0.590(5)\pi]$ & 7.842(8)& 0.10 \\
\hline
$[0.0620(6)\pi,0.557(6)\pi]$ & 7.840(6) & 0.08 \\
\hline
$[0.107(2)\pi,0.395(3)\pi]$ & 7.789(5) & -0.57\\
\hline
$[0.259(1)\pi,0.577(8)\pi]$ & 7.815(7) & -0.24  \\
\hline
\end{tabular} 
	\caption{
Measured microwave field strengths $\norm{\vec{\mathcal{B}}^{\mu\text{w}}}$ and relative errors $\Delta \mathcal{B}^{{\mu\text{w}}}$ \emph{w.r.t.} the mean of all measured strengths for different orientation of the bias field $(\alpha,\beta)$. The first two entries correspond to the $2$ directions for $PE_2$~\cite{Thiele_Self_2018}, the lower $4$ entries are the reference field and the $3$ probe fields measured for the vector magnetometry.}
\label{tab:intensities}
\end{table}
To quantify, it is instructive to calculate the relative error 
\begin{equation}
\Delta \mathcal{B}^{\mu\text{w}}=\frac{\norml{\vec{\mathcal{B}}^{\mu\text{w}}}-\overline{\norml{\vec{\mathcal{B}}^{\mu\text{w}}}}}{\overline{\norml{\vec{\mathcal{B}}^{\mu\text{w}}}}},
\end{equation}
with $\overline{\norml{\vec{\mathcal{B}}^{\mu\text{w}}}}$ the mean over all measured strengths.
$\Delta \mathcal{B}^{\mu\text{w}}$ varies over $\lesssim 1\%$ over all our measurements taken within $\sim24$ hours, which is larger than the $\sim0.02\%$ ($PE_2/PE_3$-measurement) to $0.3\%$ ($PE_1$-measurement) uncertainties with which we determine the amplitudes of the individual microwave-field components (see, \emph{e.g.}, Sec.~\ref{sec:Data analysis} and Tab.~\ref{tab:rabimeasurements}). This drift is not surprising, as neither the amplifier nor the transition from the waveguide to air (and the glass cell) were carefully impedance-matched, which can lead to power reflections. Furthermore, the temperature of the amplifier is not stabilized, while the temperature of the cooling water drifts by $\sim\pm2^\circ$C.


\section{Data analysis}
\label{sec:Data analysis}

In this section we provide details on how we arrive at the measured Zeeman shifts and amplitudes of the microwave field components used in~\cite{Thiele_Self_2018}. First, we discuss how we obtain the datapoints (Fig.~\ref{fig:Spectrum}). We then discuss the fitting model used to extract Zeeman shifts and Rabi rates, and how we determine the errors with bootstrapping the uncertainties. Finally, we discuss the calculation of the magnetic dipole moments that are used to convert the measured Rabi rates to microwave field components.

\subsection{Data points}
\label{sec:Data points}
To detect the presence of an atom, its fluorescence photons are collected on a CCD camera during a PG-cooling phase (PGC-image)~\cite{kaufman_cooling_2012}. We determine the population in $\ket{g}=\ket{\text{F}=1,m_\text{F}=1}$ for each trap $k\in[1,9]$ as the ratio $p_k$ of the number of times that an atom is observed in the second image and when it was loaded, \emph{i.e.} it was detected in the first image. The statistical uncertainty $\delta p_k$ of $p_k$ is then given by the standard error of the mean of a Bernoulli-distribution $\delta p_k=p_k(1-p_k)/\sqrt{N}$.

The datapoints and errors showing the (ground-state) population $P_i^{(\alpha,\beta)}$ in state $\ket{g}$ for the transition $\ket{g}\leftrightarrow \ket{e_i}$ [$i\in(\pm,\pi)$], see Fig.~\ref{fig:Spectrum} in~\cite{Thiele_Self_2018}, are calculated from $p_k$ by taking the loading-weighted mean and error over all $N=9$ traps as:
\begin{equation}
P_i^{(\alpha,\beta)} = \sum_{k=1}^N p_k \frac{w_k}{\bar{w}} \pm \sqrt{\sum_{k=1}^N \frac{\delta p_k^2}{N^2} \frac{w_k}{\bar{w}}},
\end{equation}
where $w_k$ is the number of times, that an atom was loaded into trap $k$, and $\bar{w}$ its sum over all traps.

\subsection{Measurements and simple model based on Rabi's equation}
\label{sec:RabiModel}

\begin{figure}[t] \centering \includegraphics[width=86mm]{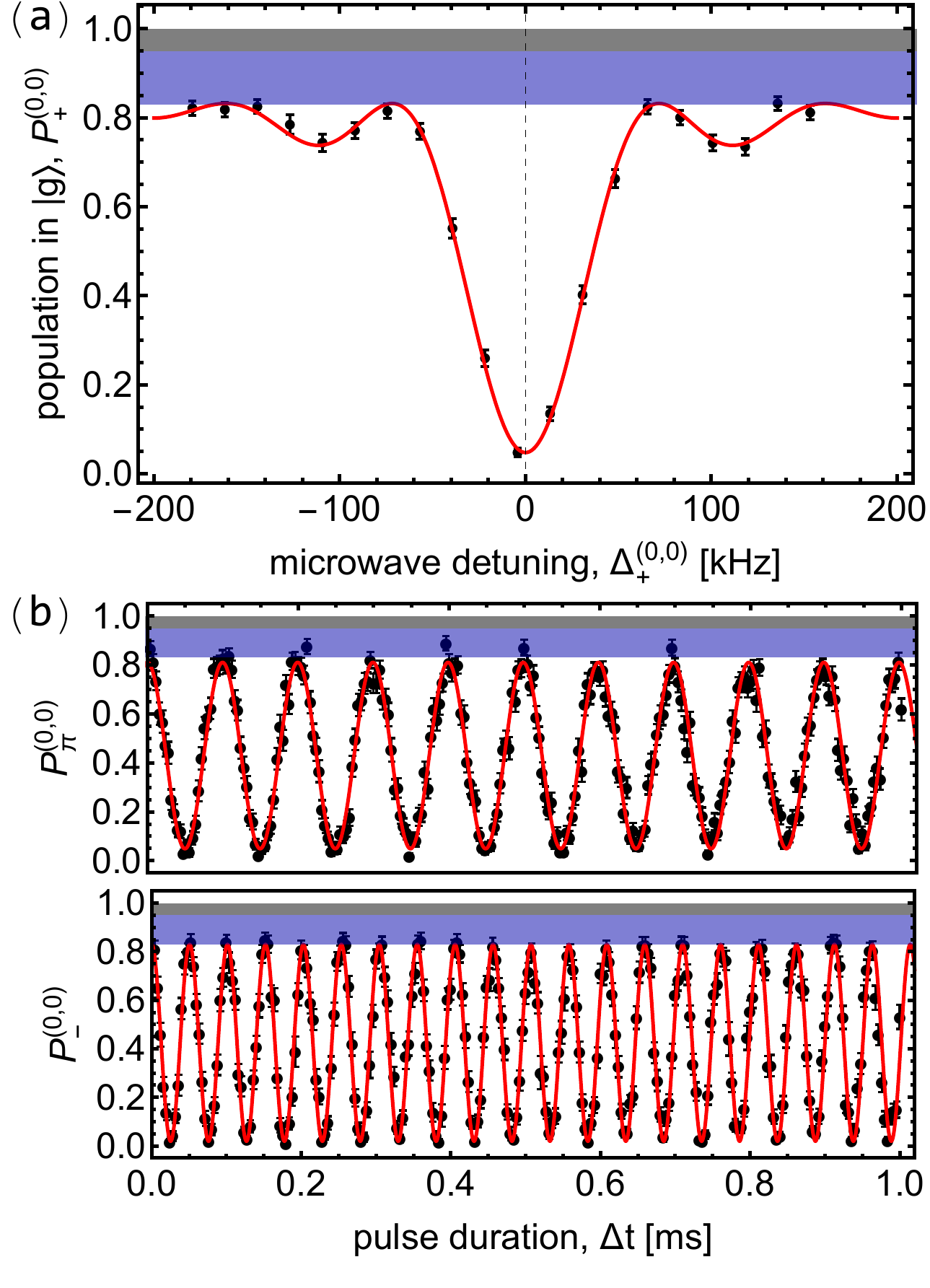}
\caption{Example measurements for different transitions $i\in(\sigma_\pm,\pi)$, when $\vec{B}^\text{bias}$ points in $\vec{z}$-direction. (a) Measured microwave spectrum (black data) obtained by varying the microwave detuning $\Delta_+^{(0,0)}$ and measuring population in $P_+^{(0,0)}$ for $\Delta t=12~\mu$s. (b) Measured coherent population transfer (black data) between $\ket{e_\pi}$ and $\ket{g}$ (upper panel) and $\ket{e_-}$ and $\ket{g}$ (lower panel).
The red lines in all plots are fits to Eq.~(\ref{eq:fitequation}) and the shaded regions indicate either heating-induced loss (blue) or loss from background collisions (black).
}
\label{fig:Spectrum}
\end{figure}

For all measurements described in the main article, we \emph{first} use pulsed coherent microwave spectroscopy to determine the atom resonance frequency $\nu_i^{(\alpha,\beta)}$ of the transition $i$ [from now on: $i \in (\sigma_\pm,\pi)$], see example in Fig.~\ref{fig:Spectrum}(a). For this, we apply a square microwave pulse of frequency $\nu$ and pulse-length $\Delta t\approx\pi/\Omega_i^{(\alpha,\beta)}$ ($\Omega_i^{(\alpha,\beta)}$ being the inverse of the transitions Rabi rate), the detuning $\Delta_i^{(\alpha,\beta)}/2\pi=\nu-\nu_i^{(\alpha,\beta)}$ of which we vary.
\emph{Second}, we set the pulse-detuning $\Delta_i^{(\alpha,\beta)}=0$ and extract the Rabi rates $\Omega_i^{(\alpha,\beta)}$ from coherent population transfer by changing the length of the microwave pulse $\Delta t$ [Fig.~\ref{fig:Spectrum}(b)]. 

The main parameters needed to determine the polarization ellipse and perform vector magnetometry~\cite{Thiele_Self_2018}, are the Zeeman-shift $\Delta^\text{zm}/2\pi=\nu_i^{(\alpha,\beta)}-\nu_0$, \emph{i.e.} the difference of the transition frequency ($\nu_i^{(\alpha,\beta)}$) and the field-free hyperfine transition frequency ($\nu_0$), and the strength of the microwave field component $\mathcal{B}_i^{(\alpha,\beta)}$ that drives the transitions $i$. To obtain small uncertainties in these quantities, we need a model that accurately describes our data (red lines in Fig.~\ref{fig:Spectrum}). 

To formulate the model, we start with Rabis equation
\begin{equation}
\label{eq:rabiequation}
P_{\ket{e_i}}(\Omega,\Delta,\Delta t)=\frac{\Omega^2}{\Omega^2+\Delta^2} \sin\left(\sqrt{\Omega^2+\Delta^2} \frac{\Delta t}{2}\right)^2,
\end{equation}
which (for a two-level system initially in its ground state) describes the population in the excited state after interaction with a square-pulse of pulse-length $\Delta t$, frequency $\nu$ (atom-pulse detuning: $\Delta$), and the Rabi rate $\Omega$. We modify Eq.~(\ref{eq:rabiequation}) taking into account that we detect $\ket{g}$, state-preparation and detection errors $A$ and $C_i^{(\alpha,\beta)}$, and a dephasing rate $\tau_i^{(\alpha,\beta)}$:
\begin{align}
\label{eq:fitequation}
&P_\text{i}^{(\alpha,\beta)}= (A - C_i^{(\alpha,\beta)})  \\ 
&\left\{e^{-\frac{t}{\tau_i^{(\alpha,\beta)}}} \left[P_{\ket{e_i}}(\Omega_i^{(\alpha,\beta)},\Delta_i^{(\alpha,\beta)},\Delta t) - 0.5\right]+ 0.5\right\} + C_i^{(\alpha,\beta)}. \nonumber
\end{align}
with the Rabi rate
\begin{equation}
\label{eq:Rabiratesuppl}
\Omega_i^{(\alpha,\beta)}=\mu_i \mathcal{B}_i^{(\alpha,\beta)}/\hbar
\end{equation}
that depends only on the amplitude of the field component $\mathcal{B}_i^{(\alpha,\beta)}$ and the transition dipole moment $\mu_i$. 

Only the $\sigma_+$-component along the coil in the $\vec{x}_\text{c}$-direction $(0,\pi/2+\delta\beta_x)$ (for definition of $\delta\beta_x$, see Sec.~\ref{sec:biasmodel}) shows a measurable decay of the Rabi contrast, probably because of polarization field gradients in the traps. The exponentially decaying contrast in Eq.~(\ref{eq:fitequation}) describes our data well and indicates that ultimately our coherence is limited by fast fluctuations of the static magnetic field strength, and not fast fluctuations of the microwave field strength that would lead to a gaussian-shaped decay of the contrast. 
\begin{table*}[t]
	\centering
  	\begin{tabular}{|l||c|c|c|c|c|c|c|}
	\hline
\boldmath$i$, $(\alpha,\beta)$ & \boldmath$A$ & \boldmath$C_i^{(\alpha,\beta)}$ & \boldmath$\tau_i^{(\alpha,\beta)} ~\text{[ms]}$ & \boldmath$\Delta^\text{zm}/2\pi~\text{[kHz]}$  & \boldmath$\Omega_{i,\text{eff}}^{(\alpha,\beta)}/2\pi~\text{[kHz]}$  &\boldmath$\mu_i$ & \boldmath$\mathcal{B}_i^{(\alpha,\beta)}~[\mu\text{T}]$ \\ 
\hhline{|=#=|=|=|=|=|=|=|}
$+$, $(0,0)$ & 0.81(2) & 0.05(1) & $\infty$ & 2314.2(5) & 44.327(9)& $\sqrt{\frac{3}{8}} (g_\text{I}-g_\text{S}) \mu_\text{b}$ & $5.1692_{-18}^{+7}$ \\
\hline 
$\pi$, $(0,0)$ & 0.810(6) & 0.049(4) & $\infty$ & 4214.2(2) &  10.031(4) &  $\sqrt{\frac{3}{16}} (-g_\text{I}+g_\text{S}) \mu_\text{b}$ & $1.6543_{-1}^{+5}$ \\
\hline 
$-$, $(0,0)$ & 0.830(7) & 0.020(4) & $\infty$ & 6113.5(3) &  19.721(4) & $\frac{1}{4} (-g_\text{I}+g_\text{S}) \mu_\text{b}$ & $5.6334_{-17}^{+7}$ \\
\hline 
$+$, $(0,\pi/2+\delta\beta_x)$ & 0.85(2) & 0.00(2) & 1.2(1) & 2215.1(4) &   25.85(1)  & $\sqrt{\frac{3}{8}} (g_\text{I}-g_\text{S}) \mu_\text{b}$ & $3.0144_{-16}^{+9}$ \\
\hline 
$\pi$, $(0,\pi/2+\delta\beta_x)$ & 0.80(1) & 0.031(7) & $\infty$ & 4147.9(3) & 31.145(6) & $\sqrt{\frac{3}{16}} (-g_\text{I}+g_\text{S}) \mu_\text{b}$ &  $5.1365_{-12}^{+8}$ \\
\hline 
$-$, $(0,\pi/2+\delta\beta_x)$ & 0.831(6) & 0.018(4) & $\infty$ & 6081.3(2) &   18.248(4)  &$\frac{1}{4} (-g_\text{I}+g_\text{S}) \mu_\text{b}$ & $5.2125_{-17}^{+7}$ \\
\hline
  	\end{tabular} 
\caption{
Summary of the fitted parameters extracted from measured spectra and coherent population transfer to determine polarization ellipses $PE_2$ and $PE_3$~\cite{Thiele_Self_2018}. For these measurements, different transitions $i \in (\pm\sigma,\pi)$ and different orientations of the bias field $(\alpha,\beta)$ were used. $\mu_i$ are the calculated transition dipole moments and $\mathcal{B}_i^{(\alpha,\beta)}$ are the microwave field components as calculated from the bootstrapped $\Omega_{i,\text{eff}}^{(\alpha,\beta)}$ and $\Delta^\text{zm}$, see text.}
\label{tab:rabimeasurements}
\end{table*}

\subsection{Extraction of correct fit parameters and uncertainties using bootstrapping}
\label{sec:bootstrapping}

In order to determine the Zeeman-shift $\Delta^\text{zm}/2\pi$ and the microwave field component $\mathcal{B}_i^{(\alpha,\beta)}$, we fit the data (Fig.~\ref{fig:Spectrum}) with the points weighted by the inverse squared of their errors. As an example, Table~\ref{tab:rabimeasurements} summarizes the values we extract from fits to the $6$ precise measurements performed for polarization ellipses $PE_2$ and $PE_3$~\cite{Thiele_Self_2018}. 

To determine the Zeeman-shift $\Delta^\text{zm}/2\pi$ from the spectra [Fig.~\ref{fig:Spectrum}(a)] is straightforward, as $\nu_i^{(\alpha,\beta)}$ can be extracted independently of other parameters. 

However, because of the atom loss in our system, we can only determine the effective Rabi rate $\Omega_{i,\text{eff}}^{(\alpha,\beta)}=\sqrt{\left(\Omega_{i}^{(\alpha,\beta)}\right)^2+\left(\Delta_i^{(\alpha,\beta)}\right)^2}$ through the measurement of coherent population transfer [Fig.~\ref{fig:Spectrum}(b)]. Because our choice to set $\Delta_i^{(\alpha,\beta)}=0$ has the uncertainty of the fitting error of $\nu_i^{(\alpha,\beta)}$, simple error propagation does not allow to determine the error in $\Omega_{i}^{(\alpha,\beta)}$. Also, as $\Omega_{i,\text{eff}}^{(\alpha,\beta)}$ is positive semi-definite, the uncertainty in $\Delta_i^{(\alpha,\beta)}$ can bias the value of $\Omega_{i}^{(\alpha,\beta)}$ toward lower values.

Hence, to determine the value and uncertainty of $\mathcal{B}_{i}^{(\alpha,\beta)}$, we bootstrap the measured data by first generating $10^5$ normal distributed values of $\Omega_{i,\text{eff}}^{(\alpha,\beta)}$ and $\Delta_{i}^{(\alpha,\beta)}$ with standard deviations given by the uncertainties extracted from the fits. From these values we calculate the asymmetric distribution of $\Omega_{i}^{(\alpha,\beta)}=\sqrt{\left(\Omega_{i,\text{eff}}^{(\alpha,\beta)}\right)^2-\left(\Delta_i^{(\alpha,\beta)}\right)^2}$ and then quote its mean and the lower and upper bounds of the $68\%$ confidence interval. The microwave field components $\mathcal{B}_i^{(\alpha,\beta)}$ are then calculated from $\Omega_{i}^{(\alpha,\beta)}$ using \begin{equation}
	\label{eq:Rabirate}
	\Omega_i^{(\alpha,\beta)}=\frac{\mu_i \mathcal{B}_i^{(\alpha,\beta)}}{\hbar},
\end{equation}
with the proper magnetic transition dipole moment $\mu_i$ [Tab.~\ref{tab:rabimeasurements}]. As expected one can observe from the asymmetric errors that the distribution is biased toward lower values.

\subsection{Derivation of magnetic transition dipole moments}
\label{sec:muDerivation}

As explained in the previous section~\ref{sec:bootstrapping}, we determine the magnetic field components $\mathcal{B}_i^{(\alpha,\beta)}$ from the corresponding Rabi rates $\Omega_i^{(\alpha,\beta)}$ using Eq.~(\ref{eq:Rabirate}). Hence, the correct calculation of the transition dipole moment $\mu_i$ is crucial for the correct determination of the microwave polarization ellipse and the results of the article. Therefore, for completeness, we sketch their calculation in the following. 

The coupling of a microwave field $\vec{\mathcal{B}}^{\mu\text{w}}$ with rate $\omega=2\pi \nu$ and components $(\mathcal{B}_x,\mathcal{B}_y,\mathcal{B}_z)$ in the laboratory frame $\mathcal{L}$~\cite{Thiele_Self_2018}, to the hyperfine states of alkali atoms in the $^2S_{1/2}$ manifold can be described by
\begin{equation}
H_\text{c}=\mu_B(g_S \vec{\text{S}} +g_I \vec{\text{I}})\cdot\vec{\mathcal{B}}^{\mu\text{w}},
\end{equation}
where $\mu_{\text{B}}$ is the Bohr magneton, $\vec{\text{S}}$ and $\vec{\text{I}}$ are the electron and nuclear spin operators, and $g_S$ and $g_I$ their Land\'e factors. In the rotating frame of $\omega$ and using the rotating wave approximation, the Rabi frequency of the transition $\ket{F,m_{\rm F}}$ and $\ket{F',m'_{\rm F}}$ is then given by: 
\begin{equation}
\label{eq:app_general_Rabi}
\Omega=\frac{\tybra{F^\prime,m^\prime_{\rm F}}H_c\ket{F,m_{\rm F}}}{\hbar}.
\end{equation}
From this, using $\mathcal{B}_\pi=\mathcal{B}_z$, $\mathcal{B}_{\pm\sigma}=\frac{\mp \mathcal{B}_x+i\mathcal{B}_y}{\sqrt{2}}$, $S_\pm=S_x\pm i S_y$, $I_\pm=I_x\pm i I_y$ and a quantization axis chosen along the $z$-direction of the laboratory frame, one finds for the Rabi rates that drive the $\Delta m_\text{F}=m'_\text{F}-m_\text{F}=(0,\pm1)$ transitions:
\begin{align*}
|\Omega_\pi|&=\frac{\mu_{\rm B}\mathcal{B}_\pi}{2\hbar^2}\tybra{F^\prime,m_{\rm F}}(g_S S_z+g_I I_z )\ket{F,m_{\rm F}},\\ 
|\Omega_{\pm\sigma}|&=\frac{\mu_{\rm B}\mathcal{B}_{\pm\sigma}}{2\sqrt{2}\hbar^2}\tybra{F^\prime,m_{\rm F}\pm 1}(g_S S_\pm+g_I I_\pm ) \ket{F,m_{\rm F}}
\end{align*}
A basis change from $\ket{F,m_\text{F}}$ to $\ket{S,I,m_{\rm S},m_{\rm I}}$ can be undertaken, with $S_z\ket{S,m_S}=\hbar m_S\ket{S,m_S}$ and $S_\pm\ket{S,m_S}=\hbar\sqrt{S(S+1)-m_S(m_S\pm1)}\ket{S,m_S\pm1}$ (and identical definitions for $I$ and $I_z$). For the $^{87}{\rm Rb}$ ground hyperfine states, $I=3/2$ and $S=1/2$, where $\vec{F}=\vec{I}+\vec{S}$ ($\vec{L}=0$). Thus, by using Clebsch-Gordon coefficients $c_{\{m_I,m_S\}}$ to express $\ket{F^\prime,m^\prime_{\rm F}}=\sum c^\prime_{\{m_I,m_S\}}\ket{m_I,m_S}$ and $\ket{F,m_{\rm F}}=\sum c_{\{m_I,m_S\}}\ket{m_I,m_S}$, 
\begin{align*}
\mu_\pi&=\frac{\hbar |\Omega_\pi|}{\mathcal{B}_\pi} \\
&=\frac{\mu_{\rm B}}{2}\sum_{m_I,m_S}c^{\prime *}_{\{m_I,m_S\}}c_{\{m_I,m_S\}}(g_S m_S+g_I m_I ),\\
\mu_{\pm\sigma}&=\frac{\hbar |\Omega_{\pm\sigma}|}{\mathcal{B}_{\pm\sigma}}\\
&=\frac{\mu_{\rm B}}{2\sqrt{2}}\sum_{m_I,m_S}\big(c^{\prime *}_{\{m_I,m_S\pm1\}}c_{\{m_I,m_S\}}g_S m_S\\
&~~~~~~~~~~~~~~~~~~~+c^{\prime *}_{\{m_I+1\pm1,m_S\}}c_{\{m_I,m_S\}} g_I m_I \big).	
\end{align*}
The relevant transition dipole moments for our measurements are displayed in Tab.~\ref{tab:rabimeasurements}, when using $g_S=2.0023$ and $g_I=-0.000995141$~\cite{steck_rubidium_nodate}.

\section{Implementation and verification of self-calibration using microwave polarization ellipses}
\label{sec:implementation_selfcalibration}

In this section we provide additional details for the self-calibration procedure~\cite{Thiele_Self_2018}. Specifically, we first discuss a model that links the direction of a bias-field $(\alpha,\beta)$ to the experimental unknowns $\{U_j\}$, $j\in(1,N_\text{u}=9)$, \emph{i.e.} $(\alpha,\beta)=(\alpha(\{U_j\}),\beta(\{U_j\}))$. We then show how we (self-)calibrate the polarization ellipse parameters $(\mathcal{B}_x,\mathcal{B}_y,\mathcal{B}_z,\phi_x,\phi_y)$ and the $\{U_j\}$ by quadratic minimization of a set of equations. Finally, we verify our understanding of the bias-field and polarization ellipses by comparison with measurements, with specific focus on the self-calibrating aspect of the procedure.

\subsection{Model of $\vec{B}^{\text{bias}}$ using experimental unknowns $\{U_j\}$}
\label{sec:biasmodel}

In our experiments, we work with a specific model for $\vec{B}^{\text{bias}}$ in the lab frame $\mathcal{L}$, that includes our control parameters (magnetic-field coil-currents) and additional experimental sources characterized by the parameters $\{U_j\}$~\cite{Thiele_Self_2018}. The three main sources we identified in our experimental setup that change $\vec{B}^{\text{bias}}$ in addition to the applied coil-currents are external fields, current-to-field calibration errors, and coil orientation. In the following, we describe how we include the $\{U_j\}$ in a model that expresses the total bias field $\vec{B}^{\text{bias}}=\vec{B}^{\text{coil}}+\vec{B}^{\text{ext}}$ as the sum of intentionally applied coil ($\vec{B}^{\text{coil}}$) and external ($\vec{B}^\text{ext}$) fields.

The currents in our Helmholtz coils are controlled by $6$ independent current-controlled-current-sources (one for each coil), the setpoints of which we vary in order to change the strength and direction of $\vec{B}^{\text{bias}}$. The coil field can thus be parametrized as $\vec{B}^{\text{coil}}=\normbvec{\vec{B}^{\text{coil}}}\left(\epsilon_x b_{x} \vec{x}_c+\epsilon_y b_{y} \vec{y}_c+b_{z} \vec{z}_c\right)$, with $\sum_{k\in(x,y,z)} b_{k}^2 = 1$ a parametrization of the unit sphere, and $(\epsilon_x,\epsilon_y)$ calibration correction factors near unity. Here, $\vec{k}_c$ [$k\in(x,y,z)$] are the unit vectors pointing into the direction of the field at the atoms when a total current $I_k=\alpha_k \normbvec{\vec{B}^{\text{coil}}} b_{k}$ is applied in two opposite coils. The $\vec{k}_c$ are expressed as linear combinations of the lab-frame vectors $(\vec{x},\vec{y},\vec{z})$ (see Fig.~1(a) in~\cite{Thiele_Self_2018}):
\begin{eqnarray*}
\vec{x}_c=&&\cos(\delta \beta_x)\vec{x}-\sin(\delta \beta_x) \vec{z}\\
\vec{y}_c=&&\cos(\delta \beta_y)\sin(\delta \alpha_y)\vec{x}
+\cos(\delta \beta_y) \cos(\delta \alpha_y)\vec{y}\\
&-&\sin(\delta \beta_y)\vec{z}\\
\vec{z}_c=&&\vec{z}.
\end{eqnarray*}
$(b_{x},b_{y},b_{z}) = (\sin(\phi)\sin(\theta),\cos(\phi),\sin(\phi)\cos(\theta))$ define a parametrization of the relative coil-currents, hence varying $\theta$ and $\phi$ controls the direction of $\vec{B}^\text{coil}$. By variation of the coils' setpoint, we calibrate $\alpha_k$ for each coil separately by measuring $\Delta^\text{zm}$ of $\ket{2,2}$ to extract $\normbvec{\vec{B}^\text{bias}}$ at the atoms. This calibration is only exact to $\lesssim 1.6\%$ of the applied field strength. Potential calibration errors are hence considered in the model with correction factors $\epsilon_x$ and $\epsilon_y$. A potential error in the $z$-direction ($\epsilon_z$) contributes to $\normbvec{\vec{B}^{\text{coil}}}$. 

We complete the set of unknown parameters $\{U_j\}=(\delta\beta_x,\delta\alpha_y,\delta\beta_y,B^\text{s}_x,B^\text{s}_y,B^\text{s}_z,\normbvec{\vec{B}^{\text{coil}}},\epsilon_x,\epsilon_y)$ using the components of the static and $\vec{B}^\text{coil}$-independent external fields $\vec{B}^\text{ext}=(B^\text{s}_x,B^\text{s}_y,B^\text{s}_z)$. Finally, our model of $\vec{B}^{\text{bias}}$ at the location of the atoms is given by 
\begin{align}
\label{eq:Bbias}
\vec{B}^{\text{bias}}&(\{U_j\},\theta,\phi)= \\
&\normbvec{\vec{B}^{\text{coil}}}\left(\epsilon_x b_{x} \vec{x}_c+\epsilon_y b_{y} \vec{y}_c+b_{z} \vec{z}_c\right)+(B^\text{s}_x,B^\text{s}_y,B^\text{s}_z). \nonumber
\end{align}  

The Euler angles $(\alpha,\beta)$ we quote in the main article are then simply the angular components of $\vec{B}^{\text{bias}}(\{U_j\},\theta,\phi)$ expressed in spherical coordinates, such that $\left[\alpha,\beta\right]=\left[\alpha\left(\{U_j\},\theta,\phi\right),\beta\left(\{U_j\},\theta,\phi\right)\right]$.

\subsection{Quadratic minimization procedure}
\label{sec:quadratic minimization}

As explained in the main article~\cite{Thiele_Self_2018}, we determine the polarization ellipse (PE) parameters $(\mathcal{B}_x,\mathcal{B}_y,\mathcal{B}_z,\phi_x,\phi_y)$ for different PE by \emph{first} measuring a number $N_\text{meas}$ of microwave field components $\mathcal{B}_{i,\text{meas}}^{(\alpha_l,\beta_l)}$, for $i\in(\sigma_\pm,\pi)$ and $l\in N_\text{meas}$. \emph{Second}, we solve a system of equations that is given by 
\begin{align}
&\mathcal{B}_{i,\text{meas}}^{(\alpha_l,\beta_l)}=\\
&\sqrt{f_{i,l}\left(\mathcal{B}_x,\mathcal{B}_y,\mathcal{B}_z,\phi_x,\phi_y;\alpha(\{U_j\},\theta_l,\phi_l),\beta(\{U_j\},\theta_l,\phi_l)\right)},\nonumber
\end{align}
where we simplify notation by denoting the right side of Eqs.~(2) in~\cite{Thiele_Self_2018} as $f_{i,l}$ and use $\alpha(\{U_j\},\theta_l,\phi_l)$ and $\beta(\{U_j\},\theta_l,\phi_l)$ for the direction of $\vec{B}^\text{bias}$ (see previous Sec.~\ref{sec:biasmodel}) in the $l^{\text{th}}$ measurement. In order to solve this system of equations we perform quadratic minimization by determining a minimum of the following function
\begin{equation} \label{eq:quadratic_function}
\mathcal{F}_\text{comp}=\sum_{l=1}^{N_\text{meas}} \frac{\left[\sqrt{f_{i,l}}-\mathcal{B}_{i,\text{meas}}^{(\alpha_l,\beta_l)} \right]^2}{\left(\mathcal{B}_{i,\text{meas}}^{(\alpha_l,\beta_l)}\right)^2}.
\end{equation}
The normalization by $\left(\mathcal{B}_{i,\text{meas}}^{(\alpha_l,\beta_l)}\right)^2$ ensures that all measurements are treated equally in the minimization procedure. 

To determine the polarization ellipse ($PE_1$), we form $\mathcal{F}_\text{comp}$ from $N_\text{meas}=28$ measurements of $\mathcal{B}_{\sigma_-, \text{meas}}^{(\alpha_l,\beta_l)}$. Only in this case, we minimize $\mathcal{F}=a^2 \mathcal{F}_\text{comp}+\mathcal{F}_\text{zm}$ with
\begin{equation}
\mathcal{F}_\text{zm}=\sum_{l=1}^{N_\text{meas}} \frac{\left[\normbvecl{\vec{B}^\text{bias}(\{U_j\},\theta_l,\phi_l)}-\normbvecl{\vec{B}^{\text{bias},(\alpha_l,\beta_l)}_\text{meas}} \right]^2}{\normbvecl{\vec{B}^{\text{bias},(\alpha_l,\beta_l)}_\text{meas}}},
\end{equation}
\emph{i.e.} we also take into account the Zeeman-shifts in order to properly extract the length of the fields. $a=0.2$ is the ratio of the mean measurement uncertainties of $\normbvecl{\vec{B}^{\text{bias},(\alpha_l,\beta_l)}_\text{meas}}$ and $\mathcal{B}_{i,\text{meas}}^{(\alpha_l,\beta_l)}$, and weighs the contribution of $\mathcal{F}_\text{zm}$ and $\mathcal{F}_\text{comp}$ to $\mathcal{F}$ accordingly. $\mathcal{F}$ is minimized for the $14$ parameters $(\mathcal{B}_x,\mathcal{B}_y,\mathcal{B}_z,\phi_x,\phi_y;\{U_j\})$ using a Nelder-Mead algorithm (shrinking: 0.5, contraction: 0.5, expansion: 5, reflection: 1). We constrain the solutions to within $(\mathcal{B}_x,\mathcal{B}_y,\mathcal{B}_z)\in(0,10~\mu\text{T}),~(\phi_x,\phi_y)\in(-\pi,\pi);~(\delta\beta_x,\delta\alpha_y,\delta\beta_y)\in(-0.1,0.1),~(B_x^\text{s},B_y^\text{s},B_z^\text{s})\in(-10\,\mu\text{T},10\,\mu\text{T}),~\normbvec{\vec{B}^\text{bias}}\in(280\,\mu\text{T},320\,\mu\text{T}),~(\epsilon_x,\epsilon_y)\in(-0.1,0.1)]$.

For $PE_2$ and $PE_3$, we only determine the $5$ polarization ellipse parameters $(\mathcal{B}_x,\mathcal{B}_y,\mathcal{B}_z,\phi_x,\phi_y)$ from 5 measurements. In this case, we use a Gauss-Newton method (Levenberg-Marquard) and solve the system of equations $10^4$ times with randomized start positions chosen from $(\mathcal{B}_x,\mathcal{B}_y,\mathcal{B}_z)\in(0,10\,\mu\text{T}), (\phi_x,\phi_y)\in(-\pi,\pi)$. From the solutions, we select only the physically relevant ones for which $(\mathcal{B}_x,\mathcal{B}_y,\mathcal{B}_z)\geq0$, and $(\phi_x,\phi_y)\in(-\pi,\pi)$ and remove duplicates, leaving us with $4$ different solutions with slightly different convergences. Because of the non-unique character of the trigonometric functions in Eqs.~(2) in ~\cite{Thiele_Self_2018}, these are true solutions to the system of equations. A possible way to identify the correct one is to add one additional independent measurement (in addition to the $5$ measurements) to the system of equations. This is equivalent to choose the unique solution out of the 4 that correctly describes (predicts) another measurement, \emph{i.e.}~exploiting the self-calibrating character of the microwave magnetometry. Equivalently, we just take the solution closest to $PE_1$, which was determined from more measurements.

To determine the direction $(\alpha,\beta)$ of the measured fields $\vec{B}^\text{m}_{j}$ in the \emph{vector magnetometry}, we invert Eqs.~(2) in~\cite{Thiele_Self_2018} for the components $\mathcal{B}_{\sigma_-}^{(\alpha,\beta)}$ and $\mathcal{B}_{\pi}^{(\alpha,\beta)}$ with the same Nelder-Mead algorithm as for $PE_1$. We constrain the possible solutions of $(\alpha,\beta)$ to lie in $\alpha \in (0,0.35\pi)$, and $\beta\in (0.5\pi,\pi)$ [except for direction $\vec{z}_\text{c}$: $\beta\in (0,0.5\pi)$], which corresponds to a volume around $\vec{B}^\text{ref}$ in which the magnetometer works uniquely. Again, this is necessary as Eqs.~(2) in~\cite{Thiele_Self_2018} are not unique under inversion. We will discuss strategies to identify the correct solution in the chapter on vector magnetometry (Sec.~\ref{sec:VectorMagnetometry}).

\subsection{Experimental verification of self-calibration}
\label{sec:verification_selfcalibration}

Using the model and the quadratic minimization described in the previous sections (Secs.~\ref{sec:biasmodel} and~\ref{sec:quadratic minimization}), we determined values for the 9 unknown experimental parameters $\{U_j\}$. Most of these parameters describe setup imperfections for which we have limited possibilities to verify their values with separate measurements. However, in the following, we benchmark these values.

More specifically, we find $(U_1,U_2,U_3)=(\delta\beta_x,\delta\alpha_y,\delta\beta_y)=(1.3\,\text{mrad},10.9\,\text{mrad},5.4\,\text{mrad})$, below the precision with which can measure relative angles of the coils, but a maximal deviation of $\sim0.6^{\circ}$ can easily be explained by machining tolerances. Furthermore, we find $\vec{B}^\text{ext}=(U_4,U_5,U_6)=(-6.05\,\mu\text{T},0.14\,\mu\text{T},-1.12\,\mu\text{T})$. We can only estimate the origin of these fields as we only measure the total sum of all (stray) external magnetic fields. The fact that this field points dominantly into the $x$-direction - the direction of the optical tweezers - might indicate effects of vector or tensor fields, the dominant contribution of which is expected to point along $x$. However, there are other sources such as large moving metallic parts that might contribute. Finally, $U_7=\normbvec{\vec{B}^\text{bias}}=302.0~\mu\text{T}$ and $(U_8,U_9)=(\epsilon_x,\epsilon_y)=(1.001,0.989)$. Here, the relative errors between the measured and the intended $300$-$\mu$T-large magnetic field ($0.66\%$), and the calibration factors $\epsilon_x$ ($0.1\%$) and $\epsilon_y$ ($1.1\%$), are indeed all smaller than our uncertainty of the current-to-field coil calibration.

In the following we compare the model and found parameters $\{U_j\}$ to measured values of both the length and direction of $\vec{B}^{\text{bias}}(\{U_j\},\theta,\phi)$. Note, that the direction measurements will show the self-calibrating property of this technique.

\subsubsection{Self-calibration model and $\normbvec{\vec{B}^{\text{bias}}}$}
\label{sec:bias_field_check}

\begin{figure}[t] \centering \includegraphics[width=86mm]{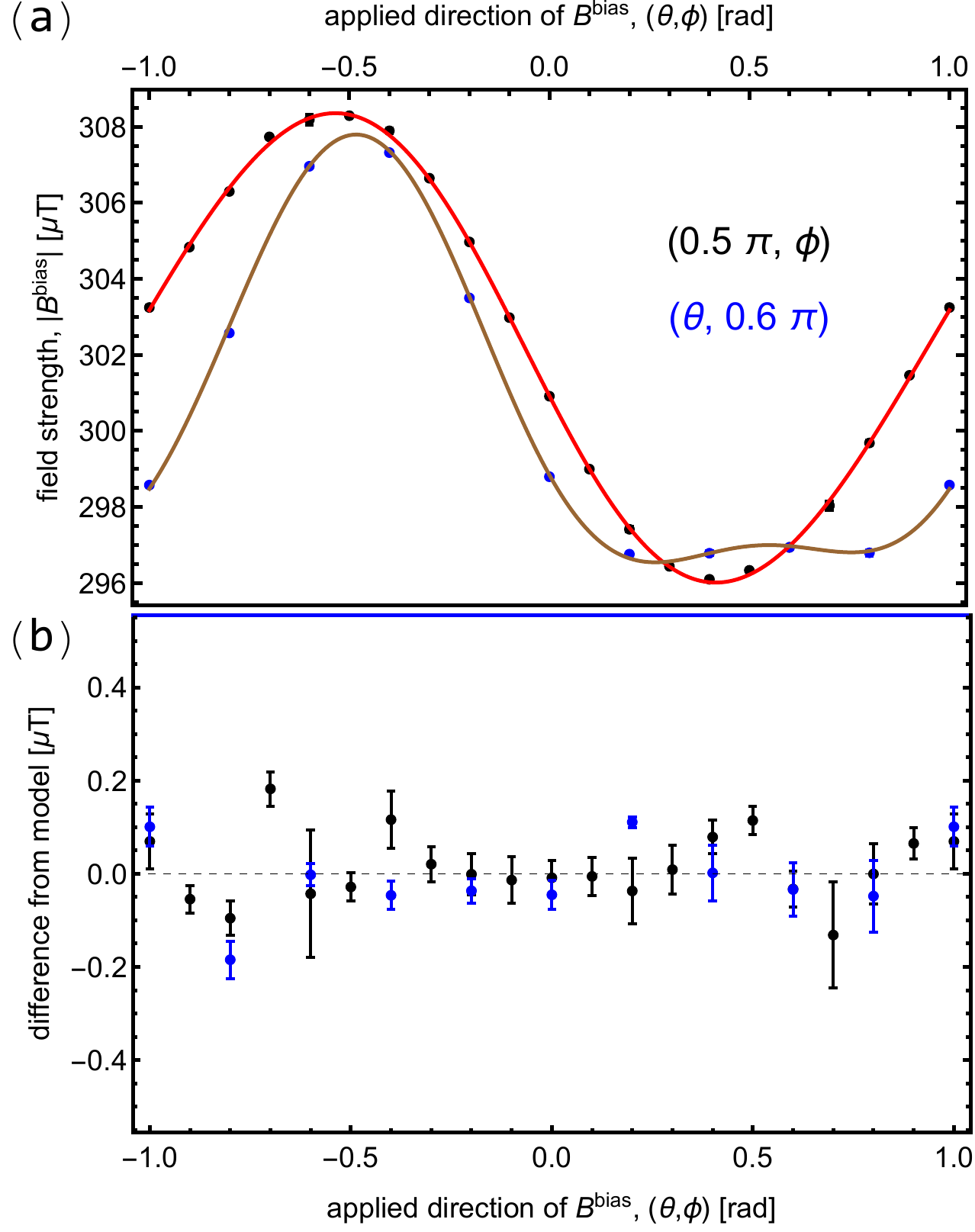}
\caption{(a) Measured bias field strength $\normbvec{\vec{B}^{\text{bias}}}$ as the direction of the field is varied experimentally. For the blue(black) data, $\theta$($\phi$) was varied for $\phi=0.6\pi$($\theta=0.5 \pi$). The brown and red lines indicate the model, Eq.~(\ref{eq:Bbias}), with parameters $\{U_j\}$ determined from the fit for $PE_1$. (b) Absolute discrepancies between measured data and model in panel (a).
}
\label{fig:FitNewAngles}
\end{figure}

In this section we compare the length of 
$\normbvec{\vec{B}^\text{bias}}$ [Eq.~(\ref{eq:Bbias})] with measurements. Fig.~\ref{fig:FitNewAngles} indicates the measured field strength $\normbvec{\vec{B}^{\text{bias}}}$ as a function of the programmatically chosen angles $(\theta,\phi)$ used to determine $PE_1$~\cite{Thiele_Self_2018}.
The red and brown lines indicate the fit (performed for $PE_1$) to the model [Eq.~(\ref{eq:Bbias})]. The small differences between the model and the measured data, indicate the accurate description of the variation of $\normbvec{\vec{B}^{\text{bias}}}$ with its direction. The remaining discrepancies of maximally $\pm0.2~\mu$T can be attributed to fluctuations in the magnetic bias-field strength.

\subsubsection{Self-calibration and microwave field component $\mathcal{B}_{\sigma_-}^{(\alpha,\beta)}$}
\label{sec:self_calibration_component}

With $\normbvec{\vec{B}^{\text{bias}}}$ in agreement with the model of and (for $PE_1$) determined values of $\{U_j\}$, we now compare the direction of $\vec{B}^{\text{bias}}$ with measurements. In addition, we show the self-calibrating character of the polarization ellipse, \emph{i.e.} that knowledge of the calibrated $\{U_j\}$ helps improve predictions using a polarization ellipse, determined from an independent set of measurements. 

In the following, we consider the three polarization ellipses $PE_1$, $PE_2$, and $PE_3$. For each ellipse, we use Eq.~(2b) in~\cite{Thiele_Self_2018}, to predict the strength of the microwave field component $\mathcal{B}_-^{(\alpha,\beta)}$ for a number of bias-field directions $(\alpha,\beta)$, as shown for $PE_1$ in Fig.~2 in~\cite{Thiele_Self_2018}. In this section, we will now focus on the differences between predicted and measured values of $\mathcal{B}_-^{(\alpha,\beta)}$. Note, that for wrongly assumed $\{U_j\}$ these differences are the sum of all functionals $\Delta\mathcal{B}$ [see Fig.~1(c) in the main article] for all sources of experimental unknowns, evaluated at the measured directions $(\alpha,\beta)$.

\begin{figure}[t] \centering \includegraphics[width=86mm]{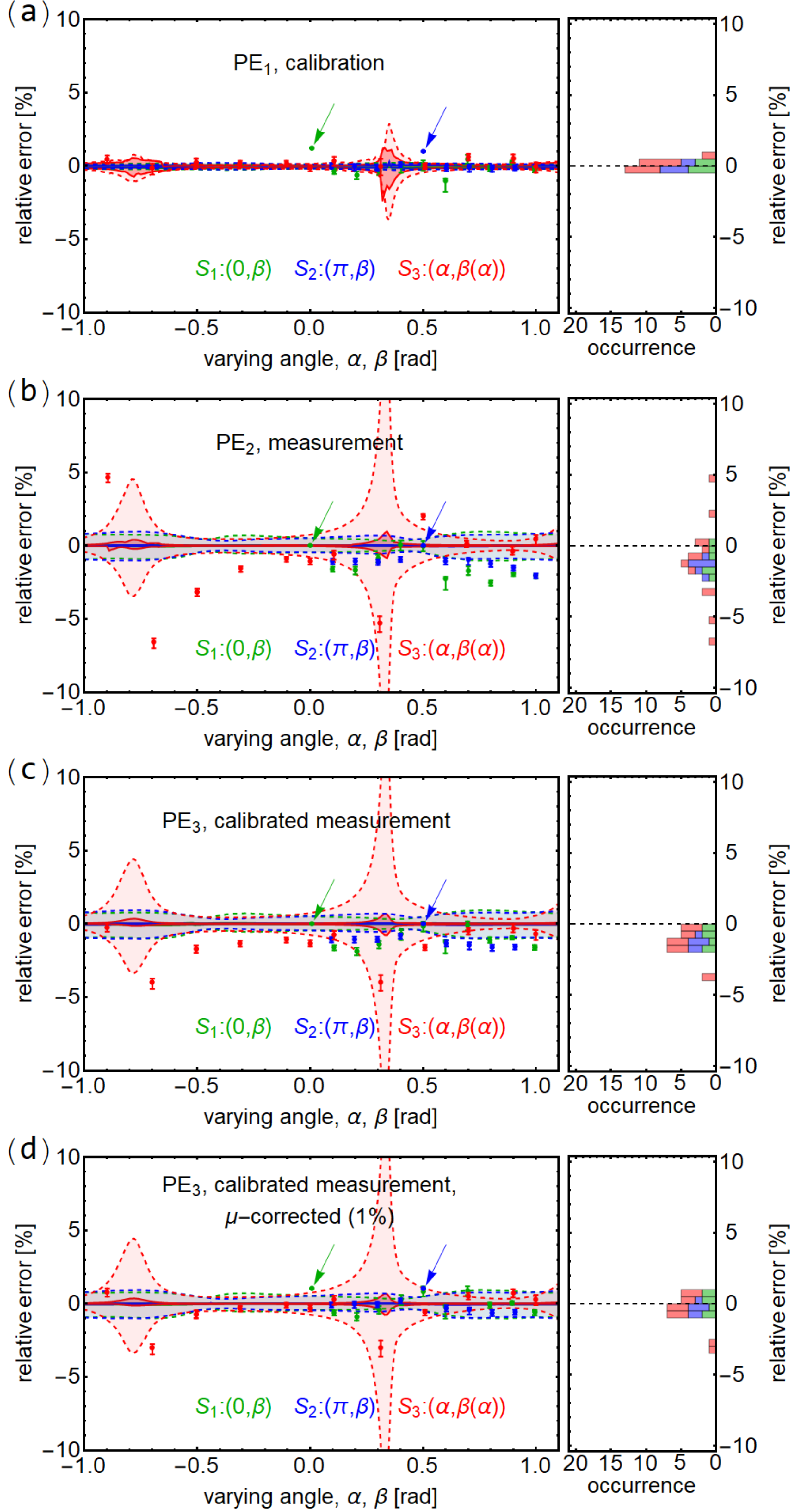}
\caption{Relative errors of measurements of $\mathcal{B}_-^{(\alpha,\beta)}$ and the prediction for polarization ellipses $PE_1$ (a), $PE_2$ (b), $PE_3$ (c), and corrected $PE_3$ (d). The green, blue and red data points correspond to measurement series $S_1-S_3$ indicated in the panels. The points highlighted by the arrows are from $S_4$ and are not included in the histograms. The shaded areas bounded by the solid (dashed) lines indicate prediction errors based on measurement uncertainties excluding (including) an additional $1\%$-drift of the microwave field strength.}
\label{fig:RabiRateDifferences}
\end{figure}

The measurements of $\mathcal{B}_-^{(\alpha,\beta)}$ to which we compare our predictions are variations of $\vec{B}^\text{bias}$ [Eq.~(\ref{eq:Bbias})] that were used to determine $PE_1$; $(\pi,\phi),~\phi\in(-\pi,\pi)$ ($18$ equidistant measurements taken in randomized order), and $(\theta,0.6 \pi),~\theta\in(-\pi,\pi)$ ($10$ equidistant measurements taken in randomized order). When calculating the values for $\vec{B}^\text{bias}$ in the laboratory frame $(\alpha(\theta,\phi), \beta(\theta,\phi))$, we obtain three series of 9, 9 and 10 measurements each [see left panels of Fig.~\ref{fig:RabiRateDifferences}(a-d)]; series $S_1$ (green data): $\sim(0,\beta),~\beta\in(0,\pi)$, series $S_2$ (blue data): $\sim(\pi,\beta),~\beta\in(0,\pi)$, and series $S_3$ (red datapoints): $\sim(\alpha,\beta(\alpha)),~\alpha\in(-\pi,\pi)$ and with $\beta(\alpha)\approx0.5\pi+0.1 \pi \cos(\alpha)$. We also show the two $\mathcal{B}_-^{(\alpha,\beta)}$ measurements used to determine $PE_2$ and $PE_3$ which are indicated by arrows in Fig.~\ref{fig:RabiRateDifferences}. Note that these measurements were taken with a $6$ times better precision than measurements in $S_1-S_3$.

In panel (a) of Fig.~\ref{fig:RabiRateDifferences}, we compare measurement series $S_1-S_3$ to the predictions of $PE_1$. The left panel shows the relative difference between the measured and predicted values of all three series. The solid lines indicate the $68\%$-confidence interval of the prediction, based on the measurement uncertainties. The dashed lines indicate the $68\%$-confidence interval of our prediction when taking into account an additional $1\%$-variation of the microwave field strength. Not surprisingly, except for one measurement of $S_1$, all measurements agree with our model as they were used to fit the model parameters. Note again, that the two measurements indicated by arrows are not part of $S_1$, $S_2$, or $S_3$, and thus were not used to determine $PE_1$. These are not in agreement with our model uncertainties, because of a change in the microwave field strength of $1\%$ between the measurement sets, see below. The stacked histogram in the right panel is identical to the (green) one in Fig.~3(c) in the main article~\cite{Thiele_Self_2018}, now colour-resolved for the three different measurement series.

To determine $PE_2$ we used measurements from series $S_4$: $\mathcal{B}_{\sigma_+}^{(0,0)},~\mathcal{B}_{\sigma_-}^{(0,0)},~\mathcal{B}_{\sigma_+}^{(0,\pi/2+\delta\beta_x)},~\mathcal{B}_\pi^{(0,\pi/2+\delta\beta_x)}$ and $\mathcal{B}_{\sigma_-}^{(0,\pi/2+\delta\beta_x)}$, and assumed \emph{wrongly} that all $U_j=0$ (except $U_7=300~\mu$T). First, we see that the measurements of $S_4$ [arrows in left panel of Fig.~\ref{fig:RabiRateDifferences}(b)] are perfectly described by the model (relative error is zero), a consequence of the $5$ performed measurements being the minimal number needed to determine a polarization ellipse. However, the (relative) differences between the predictions and (now) independent measurements of $S_1$, $S_2$, and $S_3$ span a range up larger than $\pm5\%$; discrepancies that cannot be explained, even when taking into account uncertainties because of microwave strength fluctuations or drifts (dashed lines), see Sec.~\ref{sec:microwavedrift}. As mentioned before, this is a direct consequence of the wrongly assumed experimental parameters $\{U_j\}$, \emph{\emph{i.e.}} it constitutes the sum of all functionals of $\{U_j\}$ (see $\Delta\mathcal{B}$ of Fig.~1(c) in~\cite{Thiele_Self_2018}), evaluated at the directions probed in $S_1$, $S_2$, and $S_3$. These differences are specifically pronounced for the measurements of $S_3$ (red data), which were taken out of the $(x,z)$-plane in which directions of $S_4$ lie. 

Because of the self-calibrating character of the PE, we expect that these relative errors disappear when comparing the measurements to $PE_3$, a polarization ellipse we determined using the $S_4$-series measurements \emph{and included} the correct (calibrated) values of $\{U_j\}$. 

Indeed, the relative errors between the prediction of $PE_3$ and $S_1-S_3$ now all collapse onto a band between $(-2\%,0)$ (except for 2 measurements) [Fig.~\ref{fig:RabiRateDifferences}(c)]. Specifically, this is true for the measurements of $S_3$ (red data). The systematic $1\%$-offset between the measurements and predictions can be explained by a long-term fluctuation of the microwave field strength supported by the general $\sim\pm1\%$ drifts we observe in our microwave field strength (see Sec.~\ref{sec:microwavedrift}). To account for this fluctuation, we multiply only the amplitudes $(\mathcal{B}_x,\mathcal{B}_y,\mathcal{B}_z)$ of $PE_3$ by $1.01$ [Fig.~\ref{fig:RabiRateDifferences}(d)]. The microwave-fluctuation-corrected ($\mu$-corrected) polarization ellipse $PE_3$ now predicts almost all measurements of $S_1-S_3$ correctly within the uncertainties of our model -  the self-calibration works.

\section{Details on vector magnetometry}
\label{sec:VectorMagnetometry}

In this section, we discuss additional technical details of the vector magnetometry experiment described in the main text~\cite{Thiele_Self_2018}. With our proposed technique, one can independently measure the direction and length of a magnetic bias field $\vec{B}^\text{m}$ by measuring microwave spectra and resonant Rabi oscillations (demonstrated in ~\cite{Thiele_Self_2018} using transitions $\ket{g} \leftrightarrow (\ket{e_-},\ket{e_\pi})$, respectively). Remember, in our technique we always determine a static field $\vec{B}^\text{m}$, that defines the quantization axis of our atoms. 

Depending on the origin of $\vec{B}^\text{m}$, we identified two modes of operation~\cite{Thiele_Self_2018}:
\begin{enumerate}
\item{\textbf{Reference-free mode:}} Passive operation mode, where $\vec{B}^\text{m}=\vec{B}^\text{p}$ the to-be-determined probe field, i.e. the probe field determines the quantization axis of the atoms.

This mode can in principle measure very small probe fields, as long as $\normbvec{\vec{B}^\text{p}}\gg\norm{\vec{\mathcal{B}}^{\mu\text{w}}}$. The case $\normbvec{\vec{B}^\text{p}}\lesssim\norm{\vec{\mathcal{B}}^{\mu\text{w}}}$ is still under investigation.

\item{\textbf{Reference mode:}} Active operation mode, where 
$\vec{B}^\text{m}=\vec{B}^\text{ref}+\vec{B}^\text{p}$. In contrast to the reference-free mode, an additional (known) reference field  is actively applied.
\end{enumerate}

\subsection{Determination of the length of a measured vector $\normbvec{\vec{B}^\text{m}}$}
\label{sec:VectorMagnetometry_length}

We determine the length $\normbvec{\vec{B}^\text{m}}$ of a measured static field in the same way for both operation modes by using simple Zeeman-spectroscopy and extract the Zeeman-shift $\Delta^\text{zm}$ (Sec.~\ref{sec:RabiModel}). From $\Delta^\text{zm}$ we calculate $\normbvec{\vec{B}^\text{m}}$ by using basic atomic calculations: 
\begin{equation}
\label{eq:Zeeman}
\hbar \Delta^\text{zm}\left(\normbvec{\vec{B}^\text{m}}\right)=\mu_\text{B} (g_\text{F'} m_\text{\text{F'}} - g_\text{F} m_\text{\text{F}}) \normbvec{\vec{B}^\text{m}} 
\end{equation} 
with $\mu_\text{B}$ the Bohr magneton, F (F') and $m_\text{\text{F}}$ ($m_\text{\text{F'}}$) the quantum numbers of the ground (excited) state, and
\begin{align}
g_\text{F}=&g_\text{J} \frac{F(F+1)-I(I+1)+J(J+1)}{2 F(F+1)}+\\ 
&g_\text{I} \frac{F(F+1)+I(I+1)-J(J+1)}{2 F(F+1)},
\end{align} 
with $g_\text{J}=2.00233$ and $g_\text{I}=-0.00099$~\cite{steck_rubidium_nodate}.

The field-strength sensitivity in our measurements is estimated to be
\begin{equation}
\label{eq:sensitivity}
\mathcal{S}=\mathcal{P}\sqrt{t_\text{exp}}=\mathcal{P} \sqrt{R N_\text{meas} t_\text{rep}}\approx 740\frac{\text{nT}}{\sqrt{Hz}},
\end{equation}
where the average precision, $\mathcal{P}\approx 51$ nT, is given by the fitting errors of the spectra ($0.1-1.2$ kHz) and using Eq.~(\ref{eq:Zeeman}). The duration of the experiment $t_\text{exp}\approx210$ s, is given by the number of measured microwave detunings $N_\text{meas}=20$ (pulselength: $\sim15~\mu$s), the number of repetitions $R=30$, and the duration of a single repetition $t_\text{rep}\approx350$ ms.

In our experiment, $\mathcal{P}$ is ultimately limited to $\sim 10$ nT by fast static magnetic field fluctuations. The larger uncertainty ($\leq800$ nT) we state in ~\cite{Thiele_Self_2018} is given by the standard deviation of the Zeeman shift of the three transitions after division by $m_\text{F'}+1$ [$m_\text{F'}\in(0,1,2)$], respectively. They disagree to values larger than the precision of a single measurement, because of higher order corrections of Eq.~(\ref{eq:Zeeman}), which we did not take into account here~\cite{steck_rubidium_nodate}.

\subsection{Determination of the direction $(\alpha,\beta)$ of a measured vector $\vec{B}^\text{m}$}
\label{sec:VectorMagnetometry_direction}

To determine the direction of $\normbvec{\vec{B}^\text{m}}$, we measure the Rabi rates $\mathcal{B}_{\sigma_-}^{(\alpha,\beta)}$ and $\mathcal{B}_{\pi}^{(\alpha,\beta)}$ and invert Eqs.~(2) in~\cite{Thiele_Self_2018} to find $(\alpha,\beta)$ with the (calibrated) polarization ellipse parameters from $PE_1$, see~\cite{Thiele_Self_2018} and Sec.~\ref{sec:implementation_selfcalibration}. 

In the remainder of this section, we discuss how to find the direction $(\alpha,\beta)$ of $\vec{B}^\text{m}$ (Sec.~\ref{sec:VectorMagnetometry_solutions}), and how its uncertainty depends on the measurement uncertainties (Sec.~\ref{sec:VectorMagnetometry_errorrelation}). As there are differences between the two operation modes, they will be discussed separately. 

\subsubsection{Solution for the direction $(\alpha,\beta)$ of $\vec{B}^\text{m}$}
\label{sec:VectorMagnetometry_solutions}

\begin{figure}[t] \centering \includegraphics[width=86mm]{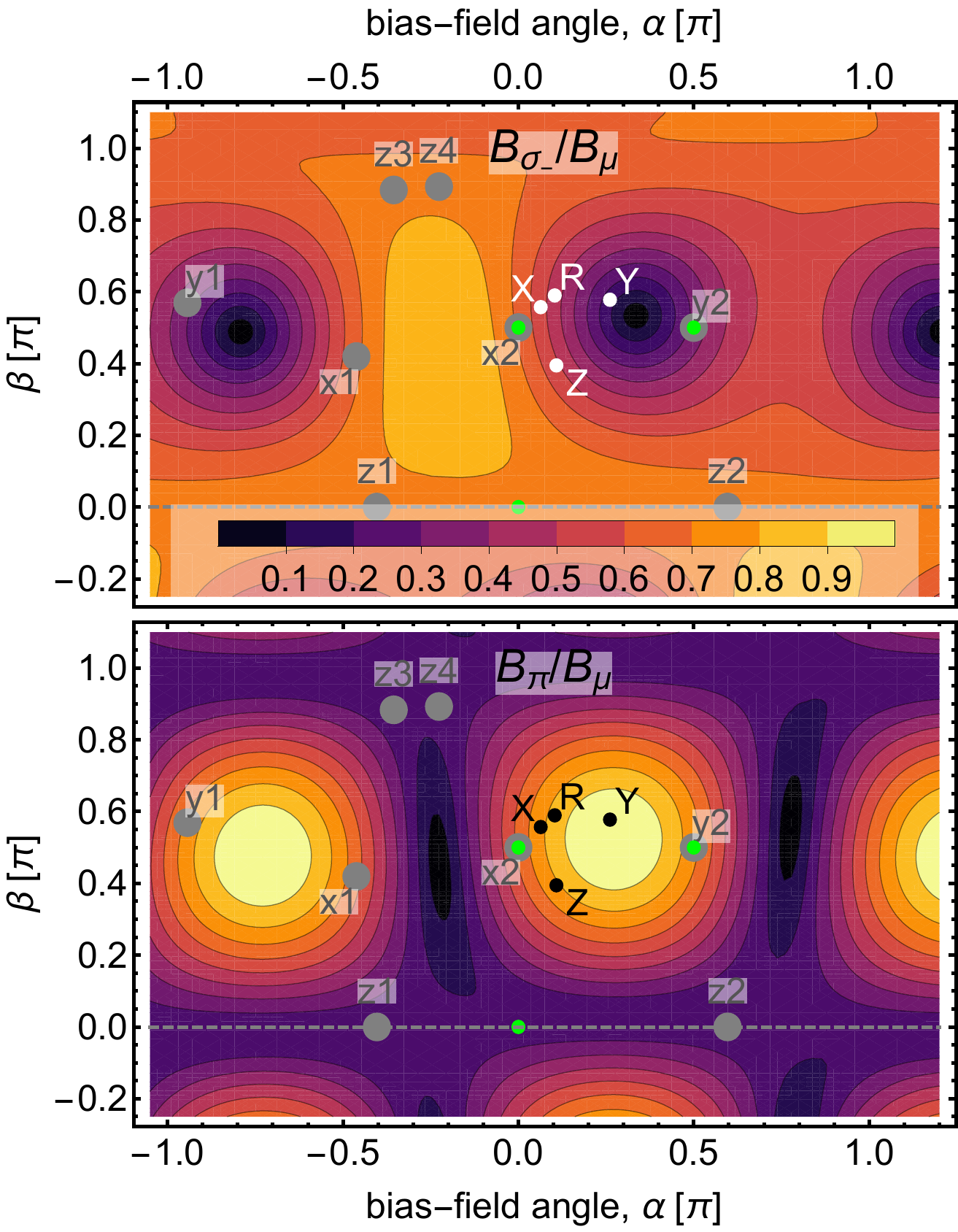}
\caption{Normalized field components $\mathcal{B}_-^{(\alpha,\beta)}$ and $\mathcal{B}_\pi^{(\alpha,\beta)}$ used to perform vector magnetometry. The components are shown for all directions of the bias field as predicted using the parameters from $PE_1$. Gray dots indicate all found solutions for the reference-free mode, whereas green dots indicate the correct directions. White/black dots indicate solutions to the reference mode, see text.}
\label{fig:FigureVecsPointing}
\end{figure}

To gain more insight in the solution-finding process, it is instructive to plot $\mathcal{B}_{\sigma_-}^{(\alpha,\beta)}$ ($\mathcal{B}_{\pi}^{(\alpha,\beta)}$) for all directions of $\vec{B}^\text{m}$ [top (bottom) panel in Fig.~\ref{fig:FigureVecsPointing}]. When solving Eqs.~(2) (main article~\cite{Thiele_Self_2018}), we cut the hyperplane of a component at the measured height $\mathcal{B}_{\sigma_-}^{(\alpha,\beta)}$($\mathcal{B}_{\pi}^{(\alpha,\beta)}$). The resulting (up to $2$) disjoint $1-$dimensional curves that accommodate the right solution, always intersect in maximally $4$ positions due to the nature of Maxwells equation. Hence, up to $4$ possible $(\alpha,\beta)$-tuplets are found that solve the system of equations, out of which the correct one needs to be identified. 

\paragraph{\textbf{Identifying the correct solution - reference-free mode:}} 

First, note that the following discussion involves measurements, that were artificially generated from $PE_1$, since we operated in the reference-mode in~\cite{Thiele_Self_2018}. In the panels of Fig.~\ref{fig:FigureVecsPointing}, we indicate all solutions found by our quadratic minimization procedure (Sec.~\ref{sec:quadratic minimization}) when applying probe fields along the $\vec{x}_\text{c}$, $\vec{y}_\text{c}$, and $\vec{z}_\text{c}$ coil directions (gray dots labeled with direction and solution number). We find $2$($4$) solutions for $x$ and $y$(z) directions, respectively. Note, that all solutions along the gray dashed line are independent of $\alpha$ (north-pole on sphere), i.e. solutions $z1$ and $z2$ are identical. 

To identify the correct direction of $\vec{B}^\text{m}$ (green dots), we propose applying a small (additional) test field $\vec{B}^\text{test}$ along an arbitrarily chosen solution. Measuring the new length of the bias field (now: $\vec{B}^\text{m}=\vec{B}^\text{p}+\vec{B}^\text{test}$), we can directly identify the correct solution as the one for which $\normbvec{\vec{B}^\text{m}}=\normbvec{\vec{B}^\text{p}+\vec{B}^\text{test}}$. 

This procedure works always, but comes with the disadvantage of an additional length-measurement and an additional test-field that needs to be applied. However, as the additional measurement only needs to discriminate between solutions, we do not expect this technique to be particular sensitive to noise in the test-field, \emph{i.e.} small $\vec{B}^\text{p}$ should still be measurable.

\paragraph{\textbf{Identifying the correct solution - reference mode:}}

To identify the correct $\vec{B}^\text{m}$, we can either use the same technique as explained in the reference-free mode, or, if we know the maximal strengths of $\vec{B}^\text{p}$, we choose $\normbvec{\vec{B}^\text{ref}}\gg\normbvec{\vec{B}^\text{p}}$. In this case, the correct solutions for $\vec{B}^\text{m}$ are the ones that are closest to the direction of $\vec{B}^\text{ref}$, as $\vec{B}^\text{p}$ only slightly perturbs the quantization axis. This is illustrated in Fig.~\ref{fig:FigureVecsPointing}, where the
direction for the reference field is indicated by the white and black points labeled 'R'. The solutions we find for $\vec{B}^\text{m}$ when applying probe fields along $\vec{x}_\text{c}$, $\vec{y}_\text{c}$, and $\vec{z}_\text{c}$ all lie in its vicinity and are close together (white/black dots labeled 'X','Y', and 'Z', respectively). This stands in contrast, \emph{e.g.}, to the maximally spaced, measured directions in the reference-free mode (green dots) of the same probe fields. 

To be precise, as in our case $\normbvec{\vec{B}^\text{p}}= {}^{2}\!/_{3} \normbvec{\vec{B}^\text{ref}}$, we approach the situation for which $\normbvec{\vec{B}^\text{ref}}\lesssim\normbvec{\vec{B}^\text{p}}$, i.e. $\normbvec{\vec{B}^\text{p}}$ can not be treated as perturbing to $\normbvec{\vec{B}^\text{ref}}$. Here, we suggest to precalculate a volume $V^{\alpha,\beta}_\text{ref,p}$ in $(\alpha,\beta)$-space around the direction of $\vec{B}^\text{ref}$ in which the correct solutions lie. Note, that $V^{\alpha,\beta}_\text{ref,p}$ depends on $\normbvec{\vec{B}^\text{p}}$ and $\vec{B}^\text{ref}$, which allows for optimization for a specific application. There may be other ways to change $V^{\alpha,\beta}_\text{ref,p}$, \emph{e.g.} by combining measurements with different $\vec{B}^\text{ref}$ or different polarization ellipses. 

\subsubsection{From measurement errors to uncertainties in solution}
\label{sec:VectorMagnetometry_errorrelation}

To judge the final performance of the vector magnetometer in terms of precision and sensitivity, one needs to assess the uncertainties in direction $(\alpha,\beta)$ of $\vec{B}^\text{m}$. These are related to the measured uncertainties of the microwave components $\mathcal{B}_i^{(\alpha,\beta)}$ in a non-trivial manner as a result of inverting ($k\in N_k\geq 2$ instances of) Eqs.~(2) in~\cite{Thiele_Self_2018}. 

To elucidate, we simplify notation by denoting the $k^\text{th}$ measurement of the microwave component $i$ as $\mathcal{B}_k^i$, and $x_j\in(\alpha,\beta)$ a choice of the to-be-determined angles. With $f^i$ the functional dependence of $\mathcal{B}_k^i$ on ($\alpha,\beta$) (see Eqs.~(2) in~\cite{Thiele_Self_2018}), we linearize the equations around their solutions $\bar{x}_j$: 
\begin{equation}
\mathcal{B}_k^i=\mathcal{B}_{0,k}^i+\sum_{j=1}^2 \frac{\partial f^i_k(x_j)}{\partial x_j}\Bigg|_{\bar{x}_j} (x_j-\bar{x}_j).
\end{equation}
Hence, the uncertainties in $\mathcal{B}_k^i$, $\Delta \mathcal{B}_k^i$ transfer to uncertainties in $x_j$ as
\begin{equation}
\Delta \mathbf{x}=\mathbf{J}^{-1}\mathbf{\Delta \mathcal{B}},
\end{equation}
where $\Delta \mathbf{x}=(x_1-\bar{x}_1,x_2-\bar{x}_2)$ are the uncertainties in $(\alpha,\beta)$, $\Delta \mathcal{B}=(\Delta \mathcal{B}_1^i,...,\Delta \mathcal{B}_{N_k}^i)$ the measurement uncertainties, and $J_{k,j}=\frac{\partial f^i_k(x_j)}{\partial x_j}\Big|_{\bar{x}_j}$ is the Jacobian Matrix. Hence, the (2-dimensional) covariance matrix $\mathbf{\Sigma}^{x}$ of the $\Delta \mathbf{x}$ relates to the ($N_k$-dimensional) covariance matrix $\mathbf{\Sigma}^{\mathcal{B}}$ of the measurements $\Delta \mathcal{B}$ via:
\begin{equation}
\mathbf{\Sigma}^{x}=\mathbf{J}^{-1} \mathbf{\Sigma}^{\mathcal{B}} \mathbf{J}^{-T}.
\end{equation}
Even for uncorrelated measurement errors ($\mathbf{\Sigma}^{\mathcal{B}}$ is diagonal), $\mathbf{\Sigma}^{x}$ is in general not diagonal. Thus, the directions of largest and smallest uncertainties (eigenvectors of $\mathbf{\Sigma}^{x}$) are not in the direction of $(\alpha,\beta)$, but in a combination of them. 

Note, one can manipulate $\mathbf{J}$ to optimize $\mathbf{\Sigma}^{x}$ for a specific application, \emph{e.g.} specific sensitivity in one direction. One strategy is to change the functional $f^i(x_j)$, \emph{e.g.} by engineering the PE parameters (possible in both operation modes). 
Another way is to change the ranges of $\bar{x}_j$, \emph{e.g.} by choosing $\vec{B}^\text{ref}$ (only possible in \emph{reference-mode}). 

\paragraph{\textbf{reference-free mode:}}
The directions of our artificially generated (see '\emph{reference-free} mode' in Sec.~\ref{sec:VectorMagnetometry_solutions}) measured fields are indicated by green dots in Fig.~\ref{fig:FigureVecsPointing}. In this case, the uncertainties can be sub-optimally large. For example, the local gradient of both microwave components at the $\vec{z}_\text{c}$-solution [$(\alpha,\beta)=(0,0)$] is small. As $\mathbf{\Sigma}^{\mathcal{B}}$ is related to $\mathbf{\Sigma}^{x}$ via the inverse of $\mathbf{J}$ [the local gradient at $\{x_j\}=(0,0)$], the uncertainties in both directions are not minimal. 

\paragraph{\textbf{reference mode:}}

We chose the direction of $\vec{B}^\text{ref}$ [$\sim(0.1\pi,0.6\pi)$, indicated by 'R' in Fig.~\ref{fig:FigureVecsPointing}] such that both microwave components have a finite gradient \emph{w.r.t.} $(\alpha,\beta)$. As a result, solutions of $\vec{B}^\text{m}$ (labeled 'X','Y','Z') also have finite gradients in both field components, leading to aforementioned assymetric uncertainties when diagonalizing $\Sigma^\text{x}$.

\begin{figure}[t] \centering \includegraphics[width=86mm]{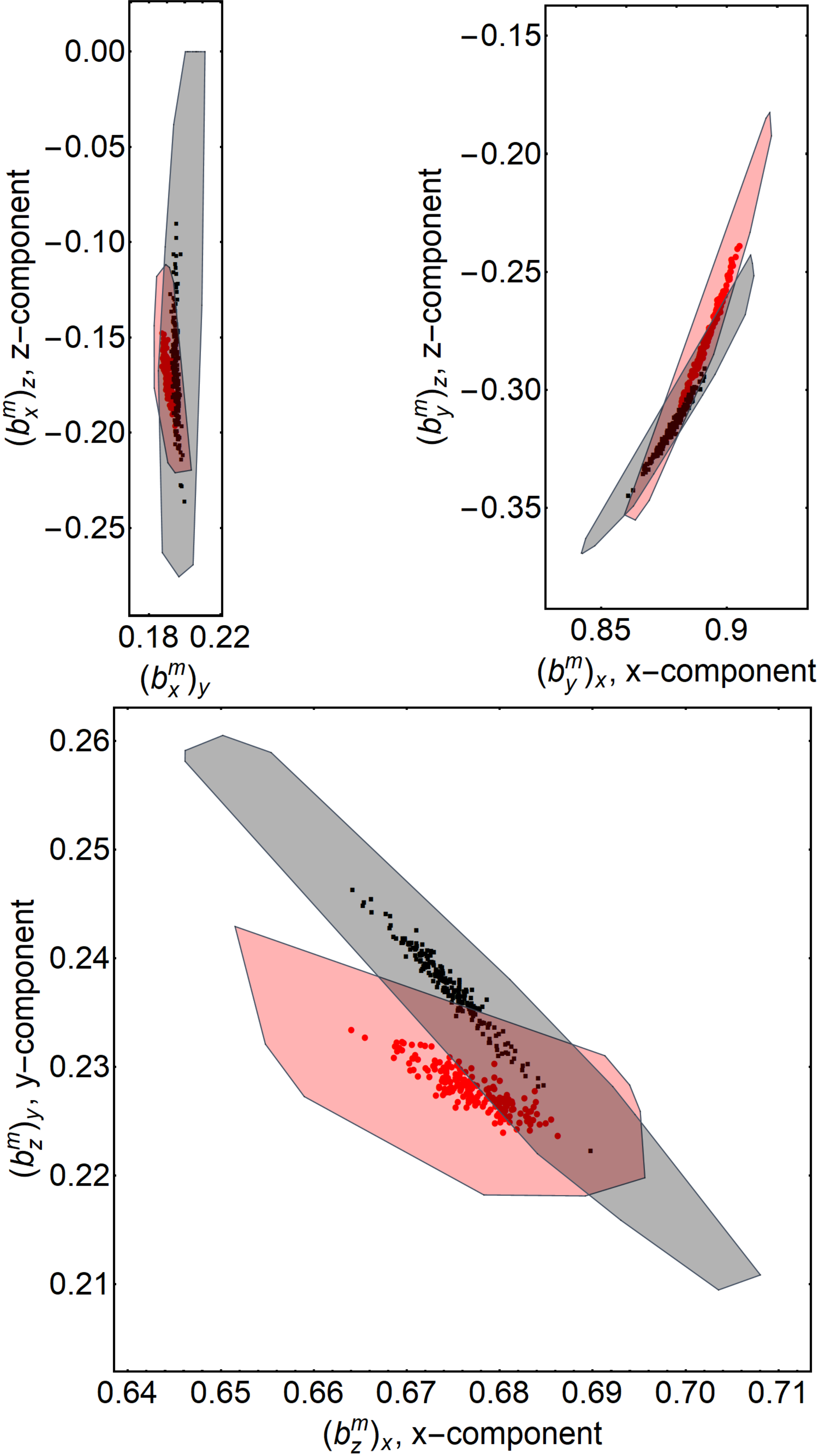}
\caption{Vector magnetometry: Field component of the length-normalized measured (black data) static field vectors taking into account the uncertainties of the measurements and the microwave polarization ellipse. The red dots indicate the expected components, based on the measurement of $\vec{B}^\text{ref}$ and our understanding of the added $\vec{B}^\text{p}$. Remaining differences can be attributed to the slow microwave-strength drifts, which are taken into account by the shaded areas, that indicate the convex hull of 200 instances of solving the system of equations. The overlap between the red and black cloud indicates that the magnetometer works within our understanding of the uncertainties. 
}
\label{fig:FigureVecPrediction}
\end{figure}
Indeed, this assymetry is observed in the measurements~\cite{Thiele_Self_2018}, when we plot two components of the normalized measured fields $\vec{b}^\text{m}=\vec{B}^\text{m}/\normbvec{\vec{B}^\text{m}}$ [Fig.~\ref{fig:FigureVecPrediction}].
For each measured field, the black dots indicate results of 200 instances for which we solved Eqs.~(2) (see~\cite{Thiele_Self_2018}), bootstrapping the uncertainties of both the measurements and of the polarization ellipse parameters. 

The precision of the different field measurement, are for the (short, long) axes: $\mathcal{P}_x\approx(0.89\,\text{mrad},10.5\,\text{mrad})$, $\mathcal{P}_y\approx(0.58\,\text{mrad},17.5\,\text{mrad})$, and $\mathcal{P}_z\approx(0.82\,\text{mrad},4.8\,\text{mrad})$ for the $\vec{x}_\text{c}$, $\vec{y}_\text{c}$, and $\vec{z}_\text{c}$ directions respectively [Fig.~\ref{fig:FigureVecPrediction}]. Considering Eq.~(\ref{eq:sensitivity}), we obtain direction-sensitivities of $\mathcal{S}_x\approx(20,243)~\text{mrad}/\sqrt{\text{Hz}}$, $\mathcal{S}_y\approx(13,401)~\text{mrad}/\sqrt{\text{Hz}}$, and $\mathcal{S}_z\approx(19,111)~\text{mrad}/\sqrt{\text{Hz}}$, taking into account $t_\text{exp}\approx525$ s ($t_\text{rep}\approx350$ ms, $R=30$, and $N_\text{meas}=50$). For completeness, we also state here the precision and sensitivities for the corresponding Rabi measurements which lie in the ranges $\mathcal{P}\approx3\text{nT}$ to $11~\text{nT}$ and $\mathcal{S}\approx 69~\text{nT}/\sqrt{\text{Hz}}$ to $252~\text{nT}/\sqrt{\text{Hz}}$.

\subsection{Vector magnetometer predicts $\vec{B}^\text{p}$}

To assess whether our magnetometer works as expected, we compare the (overlap) of the bootstrapped components of the measured $\vec{B}^\text{m}$ [black data in Fig.~\ref{fig:FigureVecPrediction}] with the expected components, calculated from the measured $\vec{B}^\text{ref}$ (and uncertainties), and adding the expected $\vec{B}^\text{p}$ (red data). Note to calculate these uncertainty ranges, both the measurement uncertainties, and the uncertainties of the $PE_1$ were taken into account. Whereas the field in $\vec{x}_\text{c}$-direction is correctly predicted, the measured and expected fields for the $\vec{y}_\text{c}$ and $\vec{z}_\text{c}$ directions agree only marginally. We attribute this to the slow drifts in the microwave field (Sec.~\ref{sec:microwavedrift}). Indeed, when taking this drift into account, all measured fields $\vec{B}^\text{m}$ agree with the predicted fields within our understanding of the errors [opaque areas in Fig.~\ref{fig:FigureVecPrediction}]. 

\subsection{Self-calibrating vector magnetometry in hot vapor cells}
\label{sec:VectorMagnetometry_precision}

The sensitivities of the experiments presented in the main article~\cite{Thiele_Self_2018} can not compete with other systems dedicated to perform high-precision magnetometry and served as proof-of-concept only. In fact, because of the techniques presented in the main article being platform-independent, we think other systems (\emph{e.g.} hot-vapor cells) are more appealing to show sensitive, high-precision measurements of both static and microwave magnetic fields. Because of this, we have not optimized either the precision or sensitivity of our measurements. 

For completeness: The largest challenges in our system are: small atom numbers ($\sim 5$ atoms) provide small signals, the choice of simple, but insensitive measurements methods (\emph{e.g.} lorentzian versus dispersive line shapes for resonance detection), and long initialization times of the magnetometer ($\sim 1750$ and $\sim23300$ times larger than the sensing (microwave pulse) time for the direction and length measurements, respectively). 

Hot-vapor cells are the current record-holder in scalar magnetometry with sensitivities at or below values of $\text{fT}/\sqrt{\text{Hz}}$~\cite{budker_sensitive_2000,allred_high_2002,kominis_subfemtotesla_2003}, see also introduction of ~\cite{Thiele_Self_2018}. Here, the strengths of hot-vapor cells such as large atom-densities leading to large optical signals, while being able to maintain sufficiently long coherence times, are promising assets. In our proposed technique, the length and direction of a static (bias) field are determined independently, and hence one 'only' needs to identify sensitive techniques to perform Rabi measurements equivalent to our Rabi measurements in a vapor-cell platform. 

In fact, sensing microwave components and field strengths in hot-vapor cells has been achieved in experiments to image microwave fields. In recent years, microwave-field detection sensitivities of $\sim1~\mu\text{T}/\sqrt{\text{Hz}}$ ($\sim12~\text{nT}/\sqrt{\text{Hz}}$ when combining all sensors) have been shown, limited by technical means such as detector speed and data saving time~\cite{Horsley_Wide_2015}. These sensitivities are already $5$ to $20$ times better than the ones shown in this experiment, such that reaching direction sensitivities $\mathcal{S}\approx0.1-1~\text{mrad}/\sqrt{\text{Hz}}$ (for the otherwise same experimental parameters) is possible. 

Although $\mathcal{S}\approx0.5~\text{mrad}/\sqrt{\text{Hz}}$ is comparable to sensitivities of recently published works~\cite{patton_all_2014}, the authors of~\cite{Horsley_Wide_2015} assess that with simple technical improvements orders of magntitude in sensitivity could be gained. With this improvements, a magnetometer based on our technique could reach (record)-sensitivities below $10~\mu\text{rad}/\sqrt{\text{Hz}}$~\cite{Afach_Highly_2015}. Furthermore, it is realistic to envision further increasing sensitivities by using techniques that continuously drive the atoms, \emph{i.e.} they do not rely on reinitialization of the atoms after readout~\cite{Gajdacz_non_2013}.

Finally, note that in contrast to other atomic vector magnetometers, our technique provides an absolute measure of the vector direction. For example, it is easy to imagine to choose calibrating the orthonormal lab frame $\mathcal{L}$ such that its $\vec{z}$-axis points along an external object, such as a magnetic coil or an excitation laser.


%

\newpage